\documentclass[manuscript,screen]{acmart}

\AtBeginDocument{%
  \providecommand\BibTeX{{%
    \normalfont B\kern-0.5em{\scshape i\kern-0.25em b}\kern-0.8em\TeX}}}

\usepackage{amsmath,amssymb,amsfonts} % amsmath amsfonts是新添加的
\usepackage{bbding}
\usepackage{subfigure}
\usepackage{mathrsfs}

\begin{document}

\title{Complicating the Social Networks for Better Storytelling: An Empirical Study of Chinese Historical Text and Novel}

\author{Chenhan Zhang}
\email{zhangch@mail.sustech.edu.cn}
\orcid{0000-0002-2352-0485}
\affiliation{%
  \institution{Southern University of Science and Technology}
  \streetaddress{1099 Xueyuan Ave.}
  \city{Shenzhen}
  \state{Guangdong}
  \postcode{518052}
}

\renewcommand{\shortauthors}{Zhang, et al.}

\begin{abstract}
Digital humanities is an important subject because it enables developments in history, literature, and films. In this paper, we perform an empirical study of a Chinese historical text, Records of the Three Kingdoms (\textit{Records}), and a historical novel of the same story, Romance of the Three Kingdoms (\textit{Romance}). We employ natural language processing techniques to extract characters and their relationships. Then, we characterize the social networks and sentiments of the main characters in the historical text and the historical novel. We find that the social network in \textit{Romance} is more complex and dynamic than that of \textit{Records}, and the influence of the main characters differs. These findings shed light on the different styles of storytelling in the two literary genres and how the historical novel complicates the social networks of characters to enrich the literariness of the story.
\end{abstract}




\maketitle

\section{Introduction}
\label{sec:intro}
Digital humanities is a transdisciplinary subject between information technologies and humanities, such as literary classics. For instance, Google makes a contribution to digital humanities by promoting the ``Google Books Library Project'' which includes millions of paper books scanned into electronic text \cite{rosati2014google}. Digital text is easier for researchers to explore than printed books, since the development of information technology has provided numerous effective tools  \cite{strehovec2016text}. In the past decade, overwhelming data science techniques have advanced the research on digital humanities; thus, components can be extracted and analyzed from literature.

A review of previous research reveals that some areas in digital humanities remain unexplored. First, mainstream studies are limited to the humanities works on the background of the Western world \cite{kirschenbaum2016digital}. It is both interesting and constructive to investigate humanities works with oriental backgrounds. Second, only a few comparative studies on literature with different styles of the same story are conducted \cite{worth2008storytelling}. Particularly, previous researches focused more on longitudinal studies, wherein researchers usually adopt a story series, such as Harry Potter Books 1–7, as the object of study \cite{chowdhury2019analysis}. A potential research interest about the same story that discovers varied features (narrative levels, characters, events) or sentiments can arise from different literature, which may be driven by literary genres or authors’ opinion, among others. Third, network study is essential for the social network of a story and any network that possesses a topological structure, which can help gain an insight into the story’s characters based on its grand narration \cite{polo2019social}. 

To fill the gap, this paper introduces a social network and sentimental analysis work on two different texts of one of the most famous Chinese story, The Three Kingdoms. In particular, we leverage the state-of-the-art natural language processing (NLP)-based model to extract the social networks in the narratives of two books. A series of descriptive statistical analysis on the extracted networks is conducted, and we discover the homogeneity and heterogeneity in terms of topological features in these networks. Additionally, we adopt the sentimental analysis to compare the evaluations on some of the main characters. The results reveal that the social network is more complicated in the narrative of the novel (\textit{Romance}) than that of the historical text (\textit{Records}). Consequently, it can be concluded that the literariness of stories has a tight relationship with the complexity of the social networks they entail.

The main contribution of this paper is as follows:
\begin{itemize}
    \item We integrate the latest NLP and network science techniques to extract and analyze the social networks of historical text and novel.
    \item We depict the difference in the dynamic social networks of the Records and the Romance, the classic historical text, and novel of the same story.
    \item A series of comprehensive case studies are performed, and we find that the historical novel complicates the social networks of characters to enrich the literariness of the story.
\end{itemize}
The remainder of this paper is organized as follows. In Section \ref{sec:lr}, the backgrounds of text mining and social network analysis researches are presented. Section \ref{sec:pre} elaborates on the network extraction approach. We perform a series of empirical studies in Section \ref{sec:result} to demonstrate the thesis of this work. Finally, this paper is concluded in Section \ref{sec:conclusion} with a summary of potential future studies.

\section{Literature Review}
\label{sec:lr}
\subsection{Social Network Analysis (SNA)}
Previous studies have demonstrated the importance of network analysis in different domains, such as complex network in supply chains \cite{bellamy2013network} and risk identification in electric industries \cite{basole2014visual}. For networks that possess a social structure, SNA can be used to study social structures by analyzing the relationships, communities, and activities through topology graph theory \cite{otte2002social,franzosi2010quantitative}. Initially, the study of SNA focuses on the network that actually exists, such as mobile social networks \cite{kayastha2011applications} (Table \ref{tab0} categorizes the metrics of SNA according to various features of social network). The development of NLP enables the extraction of the latent social network in narratives, such as literary text and news text (narrative network analysis). Recently, studies focus on narrative networks in literary works such as novels. For example, studies on Harry Potter find salient, small-world, and scale-free features in its social network \cite{waumans2015topology,zhang2014small}, and these features reveal that the story is penetrated by compact character relationships.

\begin{table}[htbp]
  \centering
  \caption{Metrics of SNA}
  \label{t1}
    \begin{tabular}{l|l}
    \toprule
    Connections & Homophily \cite{mcpherson2001birds}, Multiplexity \cite{cardillo2013emergence}, Reciprocity \cite{dobrow2012review}, Network Closure \cite{flynn2010you}, Propinquity \\
    \hline
    Distributions & Bridge, Centrality \cite{bloch2019centrality}, Density, Distance \cite{pepperberg1999rethinking}, Structural holes \cite{burt2009structural}, Tie Strength \\
    \hline
    Segmentation & Cliques, Clustering Coefficient, Cohesion \\
    \bottomrule
    \end{tabular}%
  \label{tab0}%
\end{table}%

% \subsection{Network Extraction Methods}
% Extracting characters is the most challenging problem in social network extraction. S. Sudhahar et al. presented a clear pipeline of how related techniques automate narrative network extraction \cite{sudhahar2015network}. In addition, Wauman et.al. proposed a similar network extraction method for literary texts. In Waumans’s method \cite{waumans2015topology}, authors usually narrate a story by intervening in descriptions of the conversations that occur between characters. Identifying the speakers, as well as the audience regarded as the characters in the social network, is feasible from the context of dialogs. Moreover, these conversations can be seen as interactions in the building network, in which a new edge will connect the speaker to the audience during a conversation.

\subsection{Text Mining and Natural Language Processing}
\subsubsection{Named Entity Recognition}
Named Entity Recognition (NER) is among the core tasks in NLP. In story-oriented text ming, the NER task requires that the characters and sentiment representatives are treated as entities and can be identified in the texts \cite{borrega2007we}. A bulk of computational linguistic-based NER methods are developed, which plays vital roles in NER tasks, especially the token-level tasks \cite{sang2003introduction,marrero2013named}. 

\subsubsection{Part-of-speech tagging}
Part-of-speech (POS) tagging is the process of tagging a token (a word) for a particular part of speech according to its context \cite{marquez1998part}. Table \ref{tab1} shows each type of tag with its corresponding meaning. POS tagging helps form related grammatical rules for different language patterns. 

\begin{table}[htbp]
  \centering
  \caption{POS tags \cite{Pakray2015ResourceBA}}
    \begin{tabular}{ll}
    \toprule
    Tag   & Meaning \\ \hline
    CC    & Coordinating conjunction \\
    CD    & Cardinal number \\
    DT    & Determiner \\
    EX    & Existential ÒthereÓ \\
    FW    & Foreign word \\
    IN    & Preposition of subordinating conjunction \\
    JJ    & Adjective \\
    JJR   & Adjective, comparative \\
    JJS   & Adjective, superlative \\
    LS    & List item marker \\
    MD    & Modal \\
    NN    & Noun, singular or mass \\
    NNS   & Noun, plural \\
    NNP   & Proper noun, singular \\
    NNPS  & Proper noun, plural \\
    TO    & Infinitive marker ÒtoÓ \\
    UH    & Interjection \\
    VB    & Verb, base form \\
    VBD   & Verb, past tense \\
    VBG   & Verb, gerund or present participle \\
    VBN   & Verb, past participle \\
    VBP   & Verb, non-third person singular present \\
    VBZ   & Verb, third person singular present \\
    WDT   & Wh-determiner \\
    WP    & Wh-pronoun \\
    WP\$   & Possessive wh-pronoun \\
    WRB   & Wh-adverb \\
    XNOT  & Not and nÕt \\
    \bottomrule
    \end{tabular}%
  \label{tab1}%
\end{table}%

\subsection{Deep Learning-based NLP Models}
To extract the social network of a story, characters and their connections among one another must be identified. The distribution of characters in a story is scattered and sometimes connotative. Natural language processing (NLP) technologies automate the identification of this specific information in texts, which can be a useful weapon \cite{mitri2020story}.

The popularity of deep learning has facilitated designing a number of related models to handle the subtasks of NLP, such as NER. Google proposed BERT \cite{devlin2018bert}, which substantially overcomes the limitations of existing models. BERT is based on Open AI GPT and performs attention mechanism on its model \cite{vaswani2017attention,radford2018improving}. It can predict the correct textual ID according to its entire context without a single directional limitation. In actual cases, BERT distinctly outperforms existing models in various metrics. For reference of the readers, Table \ref{tab2} compares the capacities of the most widely adopted NLP models.

\begin{table}[htbp]
  \centering
  \caption{NLP models}
    \begin{tabular}{p{11.415em}ccc}
    \toprule \\
    Model & \multicolumn{1}{p{19.585em}}{Level of long-distance semantic obtaining} & \multicolumn{1}{p{5em}}{Parallel} & \multicolumn{1}{p{6.25em}}{Bidirectional context} \\ \hline
    Word2vec \cite{mikolov2013efficient} & 1     & \checkmark  &  \\ \hline
    Unidirectional LSTM \cite{zen2015unidirectional} & 2     &  &  \\ \hline
    ELMo \cite{peters2018deep}  & 2     &  & \checkmark  \\ \hline
    OpenAI GPT \cite{radford2018improving} & 3     & \checkmark   &  \\ \hline
    \textbf{BERT}  & \textbf{3}     & \checkmark   & \checkmark  \\ 
    \bottomrule
    \end{tabular}%
  \label{tab2}%
\end{table}%

\section{Preliminary}
\label{sec:pre}
\subsection{The Story of the Three Kingdoms}
Recently, the popular video game “Total War: Three Kingdoms” captured the attention of numerous fans, and the story of the Three Kingdoms has been explained in detail to the audience. The story of the Three Kingdoms depicts the splendid and complex plot across the three kingdoms that emerged from the remnants of the Han Dynasty in the 14th century. The two most famous books based on this story, \textit{``Records of the Three Kingdoms''} and \textit{``Romance of the Three Kingdoms''}, are chosen as the research objects. The former, written by Chen Shou (B.C. 233–297), who was an official and a historian, is a biographical, historical text that chronicles the events in the three Kingdoms era by combining the respective histories of the three kingdoms. The latter is a couplet historical novel, written by the famous novelist Luo Guanzhong (B.C. 1320–1400). Its narrative is part historical, part legend, and part mythical, wherein historical facts are combined with personal opinion and folklore.

Both books’ original versions are written in classical Chinese. Text mining may lead to biases due to the complicated syntactic rules of classical Chinese and the various entities of vocabulary (i.e., similar wordings can have different meanings). In comparison, text mining in English texts is simpler, and in addition provides more straightforward intuition to non-Chinese audience. In this context, we choose a version of \textit{Romance} by CH Brewitt-Taylor \cite{brewitt1931romance} and a version of \textit{Records} by Wilt L. Idema et al. Although they are essentially similar to the original literature, differences such as the number of characters and the framework of the story indeed exist. For readers who intend to investigate the original literature, the results of this paper are for their reference only.

\subsection{SQuAD Corpus}
To train deep learning-based NLP models in supervised or semi-supervised manners, a text corpus is required as the training dataset. In this work, we employ Stanford Question Answering Dataset (SQuAD) as the text corpus. The design of SQuAD is inspired by answering questions from reading comprehension \cite{rajpurkar2016squad}. Unlike the previous datasets, the mechanism of SQuAD requires machines to select the answer from all possible candidates in the contexts rather than from a list of possible answers for each question. The answer is sometimes not a single word but a phrase, which makes the answer difficult to predict. Therefore, the robustness of models can be improved through such rigorous learning. 

In this paper, data from literary texts are limited. Therefore, we use a BERT + SQuAD method that can substantially address the problem because it can considerably improve prediction accuracy despite limited data \cite{devlin2018bert}. Furthermore, some traditional methods are still adopted in such situations.

\begin{figure}
    \centering
    \includegraphics[scale=0.4]{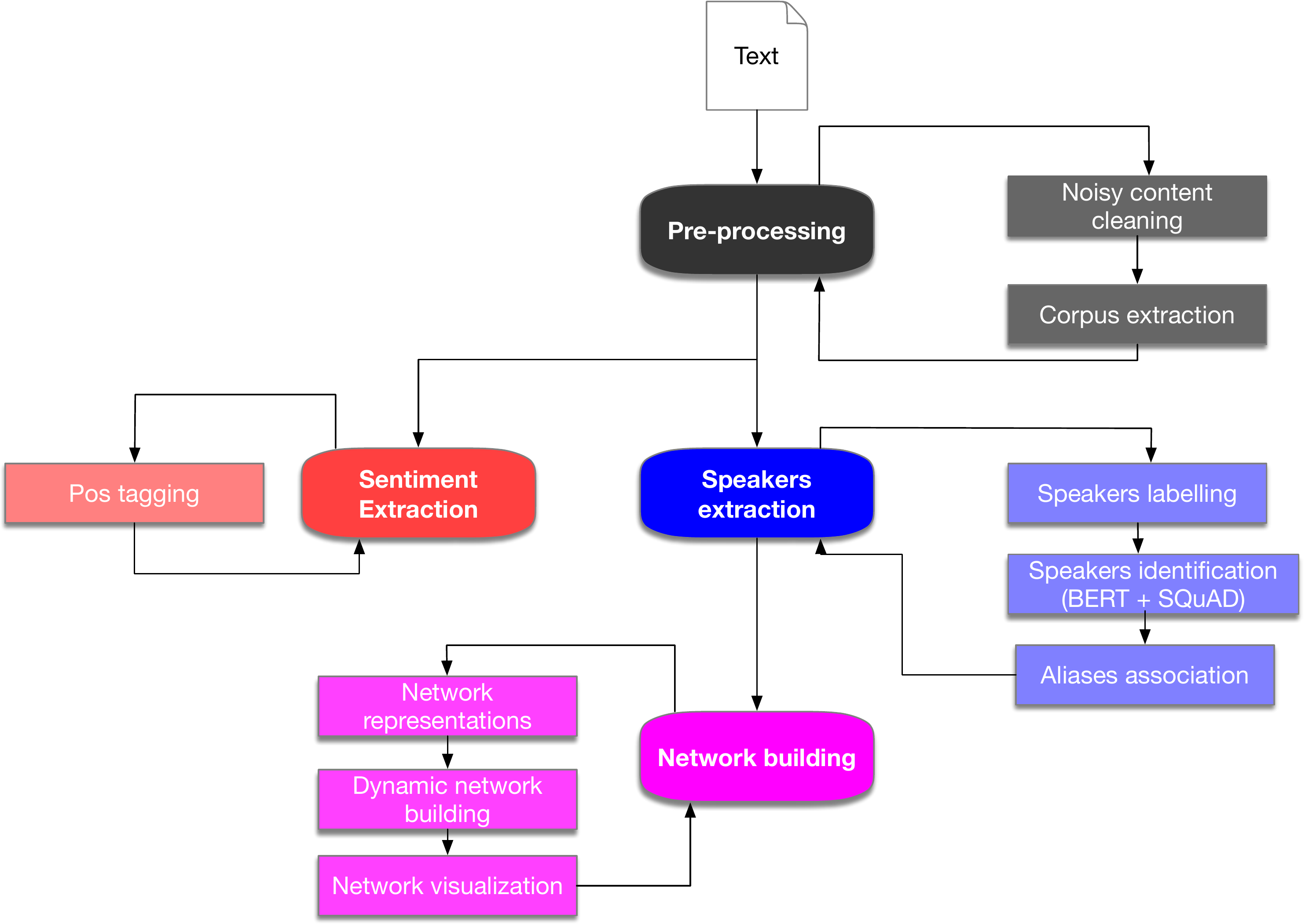}
    \caption{Overview of the entire text mining process.}
    \label{fig0}
\end{figure}

\subsection{Text Mining}
In this work, we propose a text mining algorithm to extract the social networks in the narratives. We first pre-process the raw text to clean out the noise in the text and extract the accessible text from the narrative as the corpus. Then, we identify the characters from the corpus. Meanwhile, we also achieve sentiment extraction. Finally, the extracted characters are utilized to construct the social network. The schematic view of our text mining algorithm is illustrated as Figure \ref{fig0}.

\subsubsection{Pre-processing}
Raw text is required to be cleaned and further normalized to a specific format (i.e., corpus) for the processing of the algorithm. Pre-processing work is relatively simplified in this work since the adoption of the deep learning-based tool enables a loose format of the text in the corpus.

\begin{figure}
    \centering
    \includegraphics[scale=0.4]{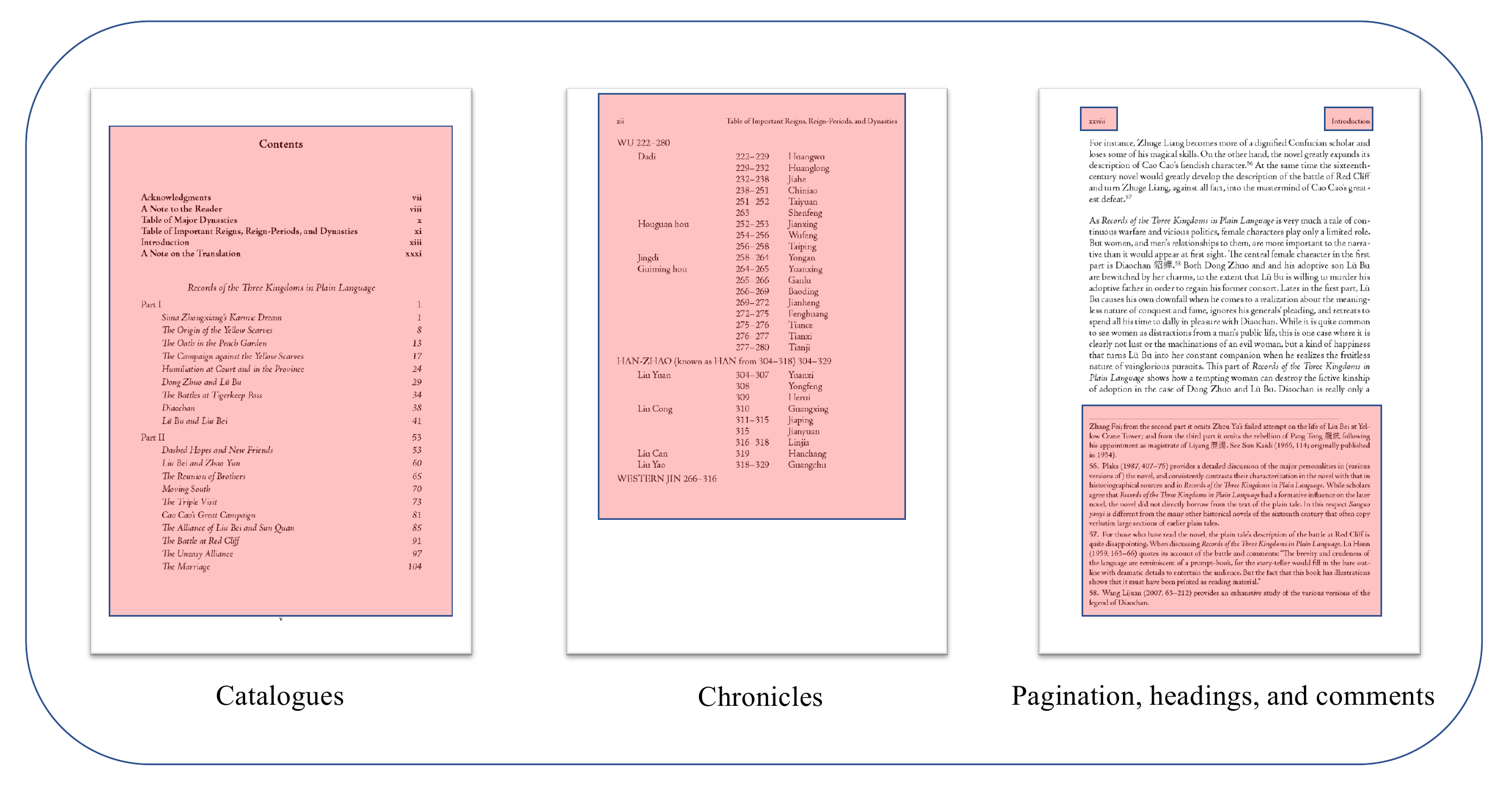}
    \caption{Noisy content in text mining.}
    \label{fig1}
\end{figure}

\paragraph{Regular expression in data cleaning}
Noisy contents are expected to be adjusted or eliminated because they are mixed with useful data, which may mislead results. Such content usually includes tables of contents, titles, headers. Fortunately, most of these noises usually follow a specific format. For example, as a translation of historical records, the “\textit{Records of the Three Kingdoms in Plain Language}” includes a multitude of notes (See Figure \ref{fig1}), where they follow the same format that starts with a serial number that leads the content. A similar phenomenon is also observed in broken words, which are all split by a hyphen or a space. Regular expressions can be used to effectively eliminate this type of noisy text by developing corresponding rules.

\begin{figure}
    \centering
    \includegraphics[scale=0.45]{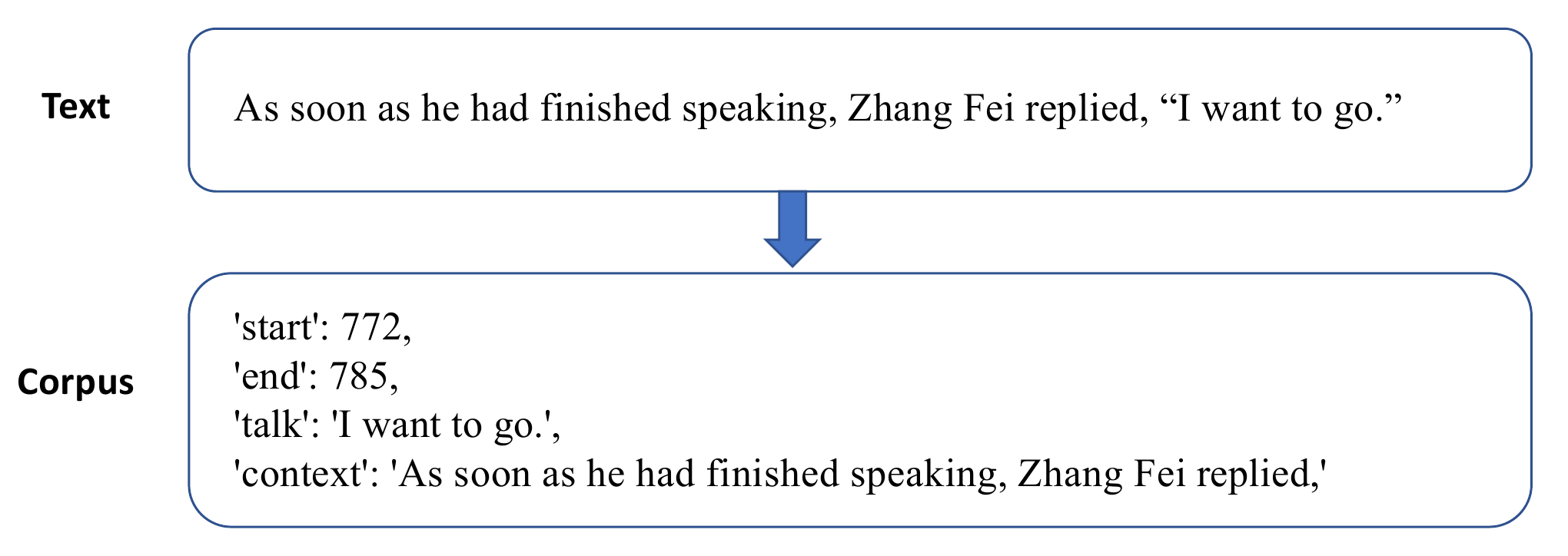}
    \caption{Mapping from the text to the corpus.}
    \label{fig2}
\end{figure}

\paragraph{Corpus Extraction}
Independently building a character-oriented corpus instead of basing on the existing corpus is essential for the character extraction task in this work. We assume that in a narrative, characters usually perform in conversations; hence, their identification is focused on such conversations. Each conversation consists of several dialogs, which usually follow a specific double quotation mark format, that is, one paragraph starts with the double opening quote (“) and ends with the double closing quote (”). Following this rule allows all dialogs to be extracted from where conversations are located. In addition, the context in which a conversation occurs commonly contains the characters (a.k.a. speakers). The context usually follows the dialog, which is easy to identify. Therefore, each item of the corpus consists of two parts, “context” and “talk,” which map to their corresponding content (See Figure \ref{fig2}).
\subsubsection{Speakers Extraction}
\begin{figure}
    \centering
    \includegraphics[scale=0.55]{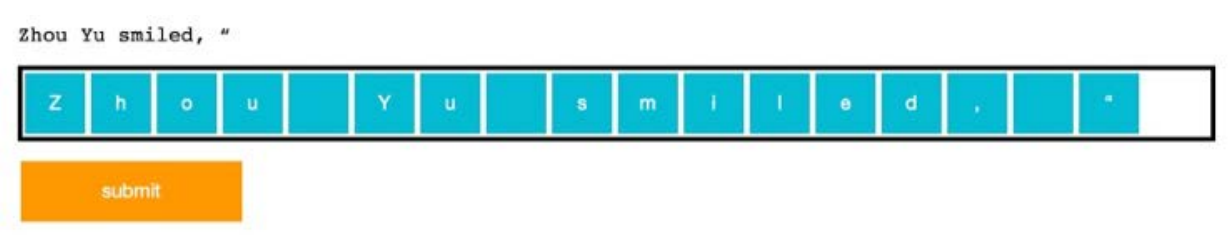}
    \caption{A visual resolution for data-labeling.}
    \label{fig3}
\end{figure}

\paragraph{Labelling the speakers}
Conversations’ portrayal varies in the storytelling. Similarly, the location of the speaker in a dialogical context differs considerably, thereby making it difficult to identify in an automated way. Therefore, a manual labeling process is required to locate the speaker in each context accurately. Given that this process is time-consuming, a GUI-labeling tool based on Jupyter Notebook is developed, and the visual operation substantially facilitates manual work (See Fig. \ref{fig3}). In this work, a total of 1,702 items from \textit{Romance} can be labeled within just three hours. 

\paragraph{Data augmentation}
The size of data extracted from books is usually insufficient to reach a promising number of training samples, and it may result that the deep learning-based models cannot achieve a satisfactory prediction result \cite{trieu2019leveraging}. In this work, the speaker corpus of Records collected only 1,248 items, and a measly portion of 806 samples (64.5\%) are labeled after transcribing the entire text. To address this issue, a data augmentation approach is introduced to generate a sufficient number of new annotated data.

How to generate new data and how much data should be generated are essential questions to answer. All speakers are assumed to be included in all the contexts. Supposing a total of $\mathcal{S}$ labelled speakers and $\mathcal{M}$ contexts, we can generate $\mathcal{D}_A = \mathcal{S}*\mathcal{M}$ new data samples. In this work, we use this data augmentation method and obtain over a million new data samples, as shown in Table \ref{tab3}.

\begin{table}[htbp]
  \centering
  \caption{Data augmentation}
    \begin{tabular}{lccl}
    \toprule
          & \multicolumn{1}{l}{$\mathcal{S}$} & \multicolumn{1}{l}{$\mathcal{M}$} &  $\mathcal{D}_A$ \\ \hline
    \textit{Romance} & 664   & 1702  & 1130128 (approx.) \\
    \textit{Records} & 806   & 1248  & 1005888 (approx.) \\ \bottomrule
    \end{tabular}%
  \label{tab3}%
\end{table}%

\begin{figure}
    \centering
    \includegraphics[scale=0.45]{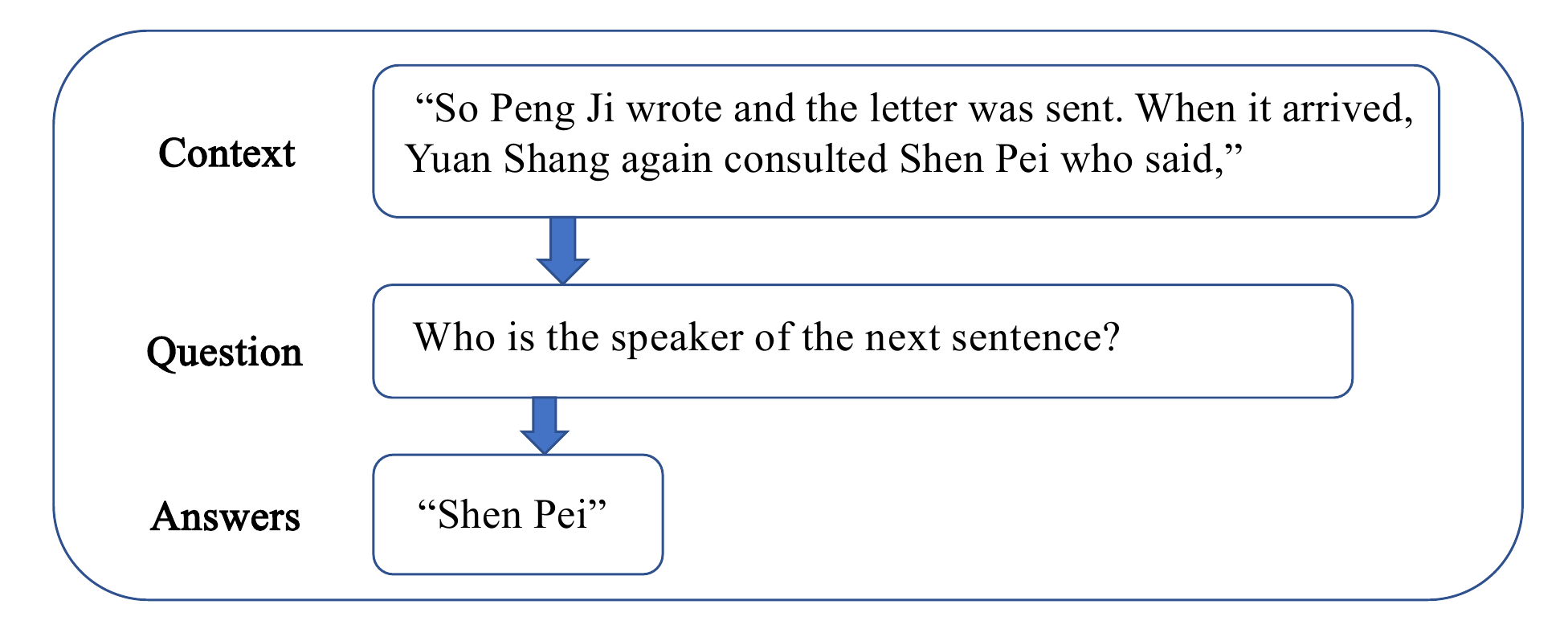}
    \caption{An example of SQuAD.}
    \label{fig4}
\end{figure}

\paragraph{Speakers Identification}
A BERT + SQuAD algorithm is used to build a speaker prediction model in this work. SQuAD provides a structure to answer the question (prediction) by comprehending the context. Referring to the structure of SQuAD, we structured a ternary dataset (i.e., context, answer, question). (See Figure \ref{fig4} for an example of the dataset).

BERT provides a contextual prediction algorithm, and we use this model to predict the speakers from the text. Specifically, we employ Google’s \verb|BERT-Base-Multilingual-Cased| model as the pre-trained model, which incorporates 12-layer, 768-hidden, 12-heads, 110 million parameters; the pre-trained model is then used to fine-tune our dataset. Note that we omit the training procedure of BERT since it is not among the main focuses, interested readers can refer to our codes available at \url{https://github.com/GreatWizard9519/Social-network-extraction-and-analysis-of-Three-kingdoms}. No related baseline evaluates the training effort because this study is an individual project. The predicted results after manual proofreading covers the vast majority of characters in the books (approximately 93\% on \textit{Romance} and 91 on \textit{Records}). In this way, the appearing characters can be obtained from the prediction result.

\paragraph{Aliases association}
More than a few characters in the two books possess one or more aliases. For example, “Xuande”, “Lord Liu”, and “The First Ruler” all refer to the character “Liu Bei”. To overcome this problem, an alias-matching mechanism is built to map aliases of the characters. A flaw of this mechanism in practical use is that the shared family name or title may be mapped to multiple characters. For example, “Sima” can be mapped to “Sima Yi” and “Sima Zhongxiang”. We develop two solutions that can solve this problem. First, the aliases mapping is classified according to the chapters of the story. For instance, “Sima Zhongxiang” is a character who simply appears at the beginning of Records; hence, the mapping: “Sima” to “Sima Zhongxiang” should solely be applied at the first few chapters. Second, the context is considered when mapping aliases. For example, when “Liu Bei” appears, the closest “Lord” should be “Lord Liu” (i.e., “Liu Bei”) with a high possibility.

\subsubsection{Sentiment Extraction}
\begin{figure}
    \centering
    \includegraphics[scale=0.5]{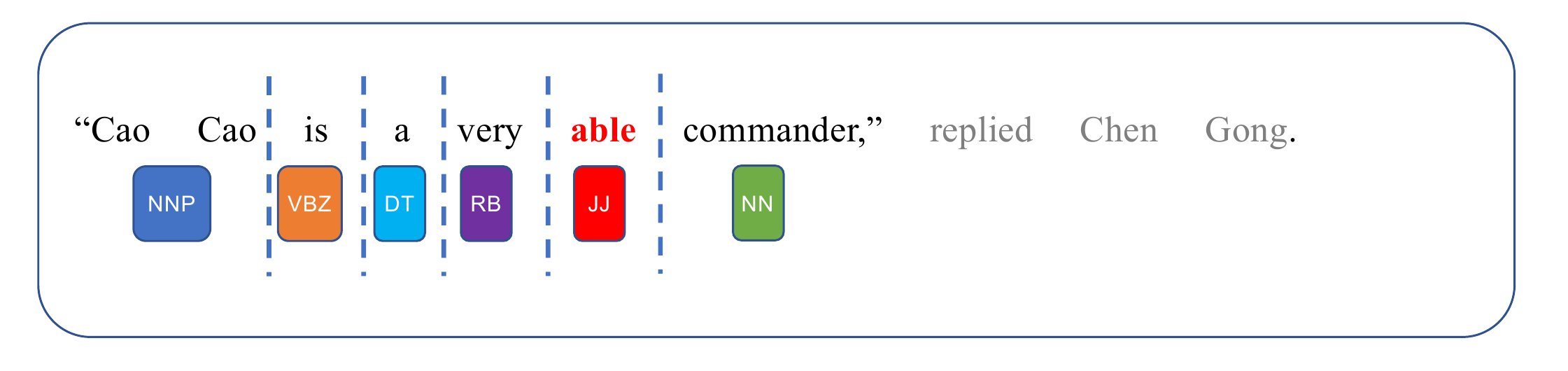}
    \caption{Sentence tokenization with POS tags.}
    \label{fig5}
\end{figure}

While the extraction and analysis w.r.t. sentiment is not a main focus of this paper, we still conduct related simple studies on some key characters to make the audience gain a deeper understanding of the story. Sentiments toward a character can be differently described. In this work, our sentiment analysis focuses on evaluative words. Other characters who comment about a certain character is a good entry to extract evaluations. Figure \ref{fig5} shows one of “Chen Gong”'s evaluations on “Cao Cao” in \textit{Romance}.

Therefore, the extraction of such evaluative words is applied. First, all conversations involving a specific character are extracted and tokenized, and each token is tagged with the corresponding part of speech (i.e., POS tag). Subsequently, following the example shown in Figure \ref{fig5}, words that possess an adjective POS (tagged with “JJ”) and collected since we consider them as ``evaluative words'' to characters, which can be utilized in sentiment analysis.

\begin{figure}
    \centering
    \includegraphics[scale=0.5]{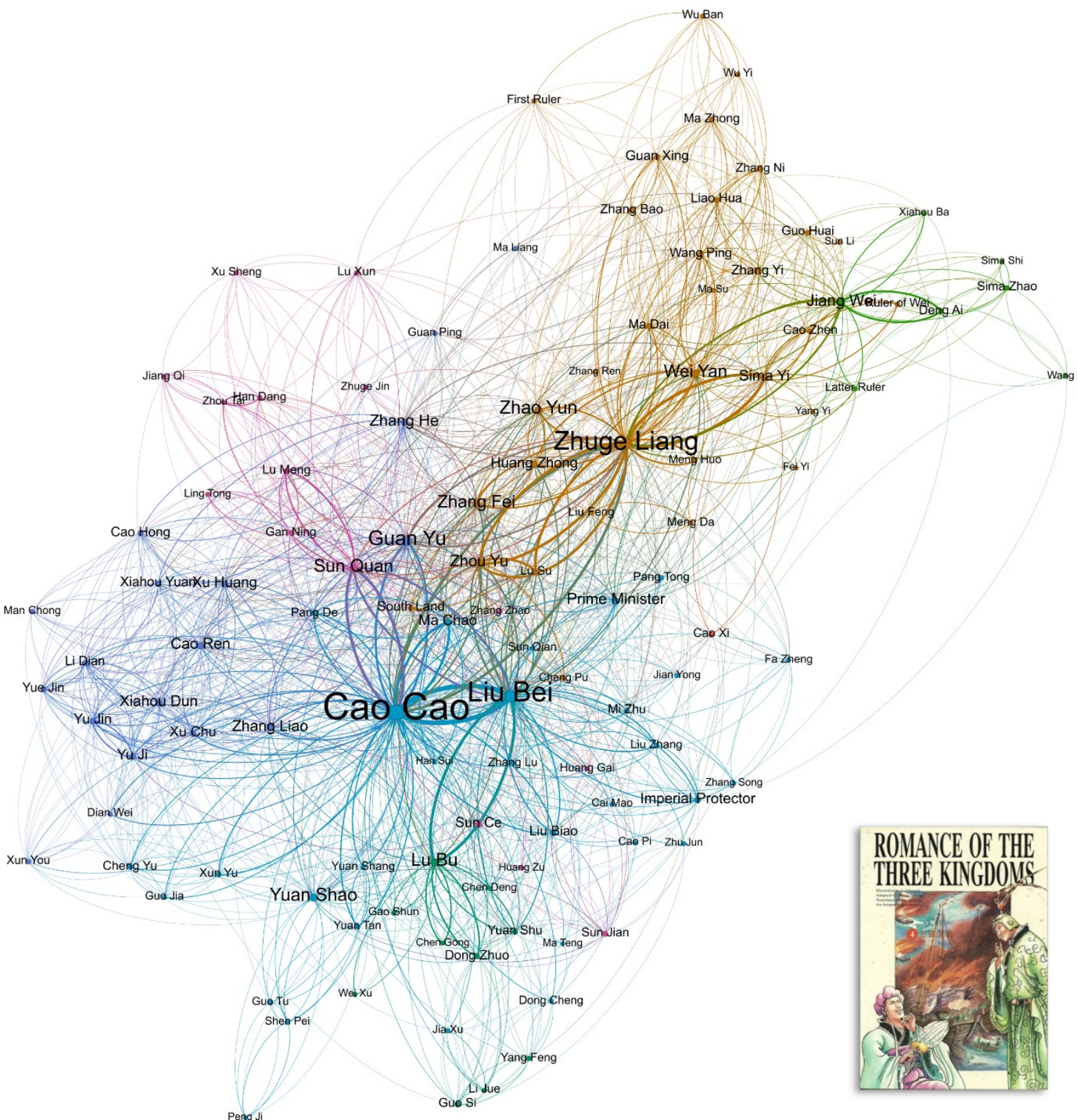}
    \caption{Network extracted from \textit{Romance} (only show nodes whose degree is larger than 6).}
    \label{fig6}
\end{figure}

\begin{figure}
    \centering
    \includegraphics[scale=0.5]{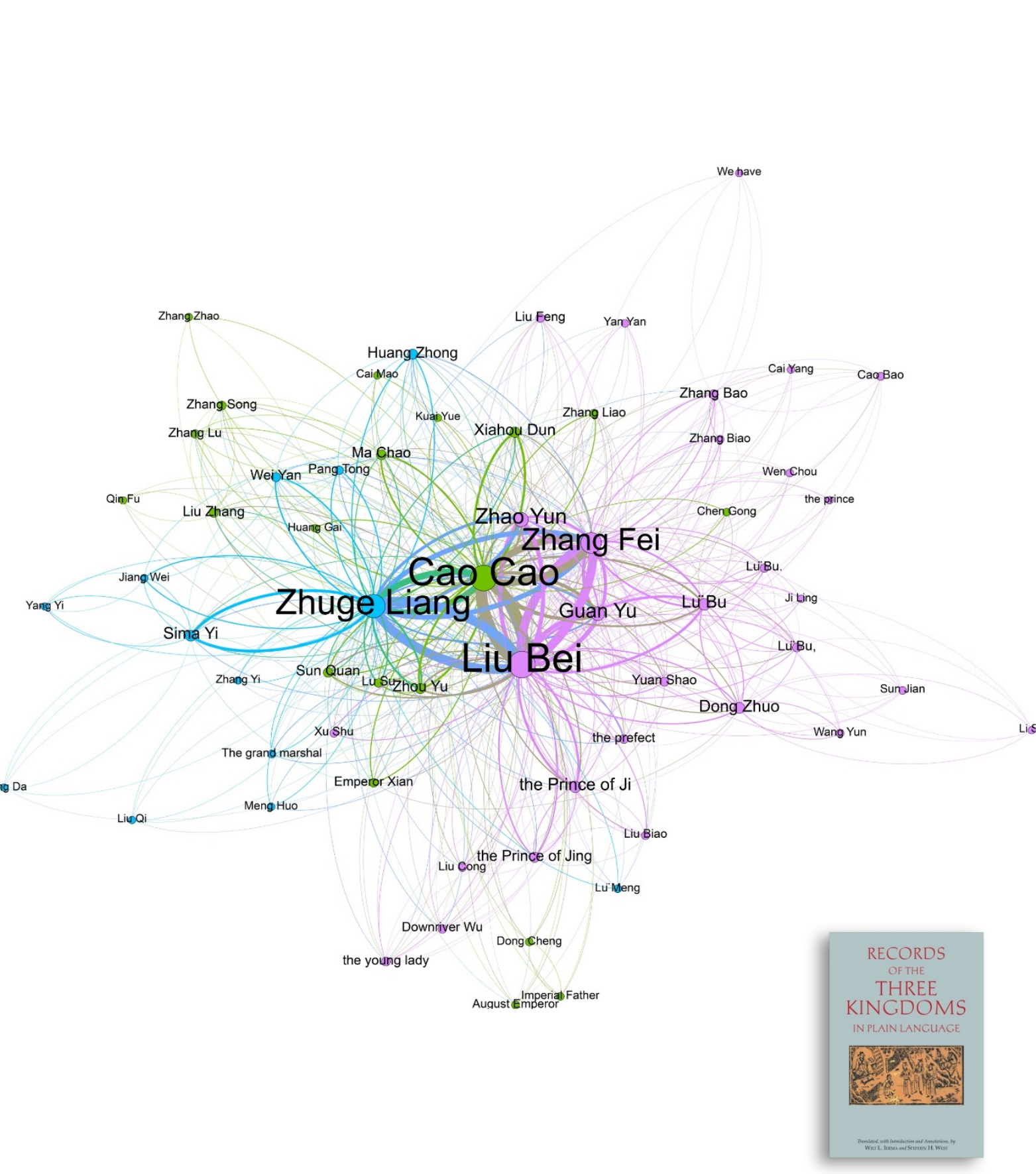}
    \caption{Network extracted from \textit{Records} (only show nodes whose degree is larger than 6).}
    \label{fig7}
\end{figure}

\subsection{Network Building}
\subsubsection{Representations}
Upon the collection of characters and the interaction that represents the nodes and edges, we can construct social networks. The essential representations for our extracted social networks are defined below:
\begin{itemize}
    \item \textbf{Nodes:} For each character coming on stage, a node is built. As aforementioned, all characters are from the identified speakers; hence, the social network merely describes the relationship between characters who have monologues or dialogs. It is worth noticing two phenomena when using this node representation. First, the number of nodes is less than the actual number of characters that appear in the books. Second, there appear some characters are isolated without any interactions with other characters (i.e., nodes whose degree is 0) in our networks. To ensure the completeness of the social network, we manually append some of the missing characters and meanwhile include the isolated nodes when constructing the networks.
   
   \item \textbf{Edge:} To correlating the nodes, namely, construct edges, we establish an assumption that the adjacent appearance of characters will serve as the basis for creating interactions. Such an assumption is seemingly a coarse-grained solution. However, the outcome will have a high degree to match the actual situation when the size of the involved data is large enough. Based on this assumption, an algorithm established that an interaction (edge) is built when two adjacent characters are detected in the same context. Furthermore, on the account that the representation of edge describes a reciprocal relationship, the network is thereby considered as a bidirectional graph, wherein the values of in-degree and out-degree of every single node are equivalent.

\end{itemize}

% \paragraph{Weight} The weight measures the number of occurrences of a character (a node) and the number of interaction that occurred between two characters (an edge).

\subsubsection{Dynamic Network}
\label{method_dynamic}
Unlike others, the social network extract from narrative will grow as the story carries on. Investigation of network dynamics can help us gain a better insight into the story. To this end, the texts of the two books are chronologically split into five stages, and their corresponding networks are extracted though the same method introduced above. Some key events are set aside as separate markers to normalize the distribution of each stage due to the difference in the chapter settings of the two books, for instance, “the death of Dong Zhuo” and “the death of Liu Bei”. Moreover, these five stages represent the five most prominent periods in the story of the Three Kingdoms. Joining the five separate networks, a dynamic network with evolving growth across the five stages is obtained.

\subsubsection{Network Visualization}
In this work, we use Gephi \cite{bastian2009gephi} to visualize the extracted social networks. To present a clear visual effect, the two demonstrated networks (See Figure \ref{fig6} for \textit{Romance} and Figure \ref{fig7} for \textit{Records}) have been filtered to only include the characters (nodes) whose degree is greater than 6. Nodes are classified by using different colors and sizes, wherein the size of nodes is ranked from its value of degree. Moreover, the color of nodes is determined by their communities categorized by modular algorithms for aesthetic needs.

\section{Results and Discussion}
\label{sec:result}
This study focuses on exploring the discrepancy or similarity of multiple dimensions between the two books of the Three Kingdoms by employing social network analysis, and further gain an insight into the storytellings entailed in the two books. Our analysis incorporates two dimensions. First, a holistic analysis on the social networks extracted from the two books is introduced wherein global properties are emphatically considered. Subsequently, the observation on some protagonists will be discussed. To present the research logic, in the following investigations, we will first raise some interesting questions and approach them with rational explanations from the analysis results.

\subsection{Global Network Analysis and Comparison}
\subsubsection{History vs Romance: Which is Grander?}
The framework of a great story is grand, which generally involves numerous characters, intricate relationships, and thus entails a vast social network. In this work, we measure the ``grandness'' of the two social networks by using the three metrics as below: 
\begin{itemize}
    \item $\mathcal{N}$: The number of characters who appear in the story (i.e., number of nodes).
    \item $\mathcal{E}$: The number of interactions that occur in the story (i.e., number of edges).
    \item $d$: The shortest distance between the two most distant nodes in the network (i.e., the diameter of network).
\end{itemize}

\begin{figure}
    \centering
    \subfigure[Top 50 in \textit{Romance}.]{
    \includegraphics[scale=0.31]{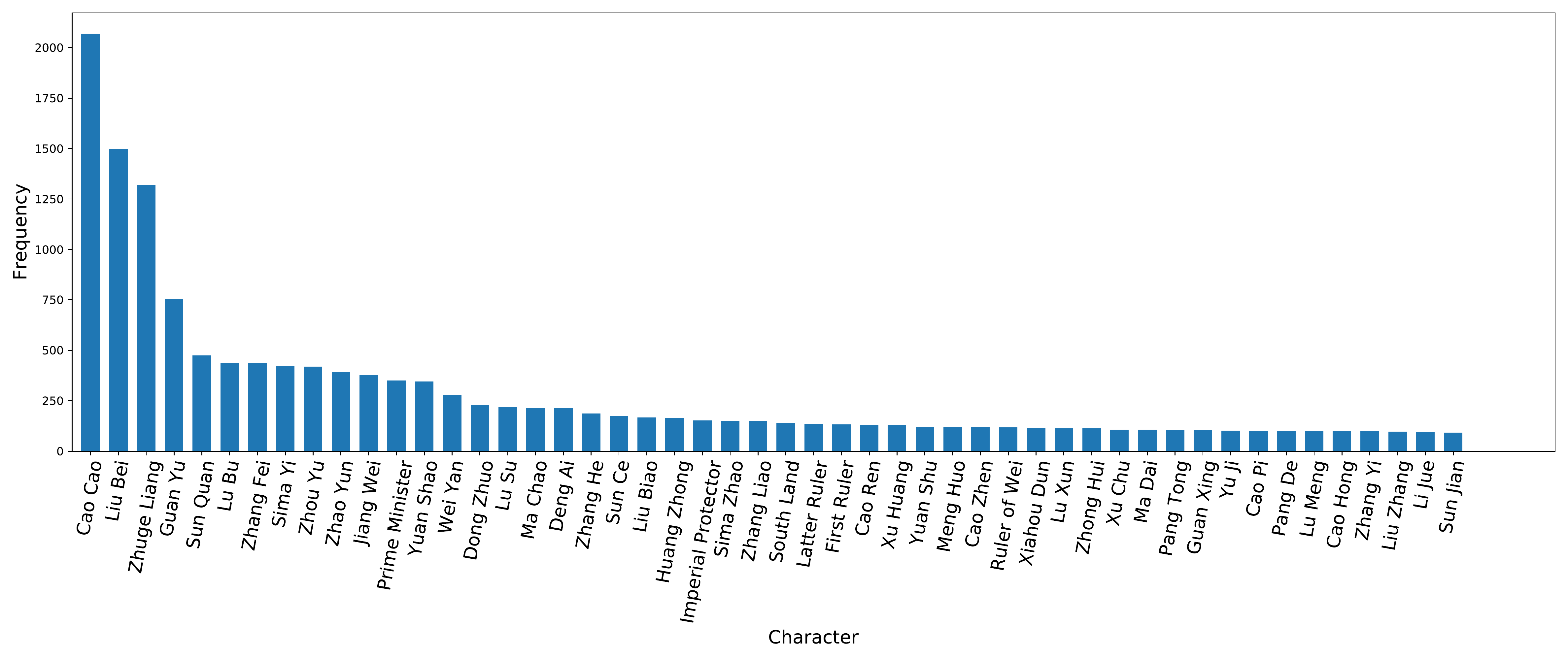}}
    \quad
    \subfigure[Top 50 in \textit{Records}.]{
    \includegraphics[scale=0.31]{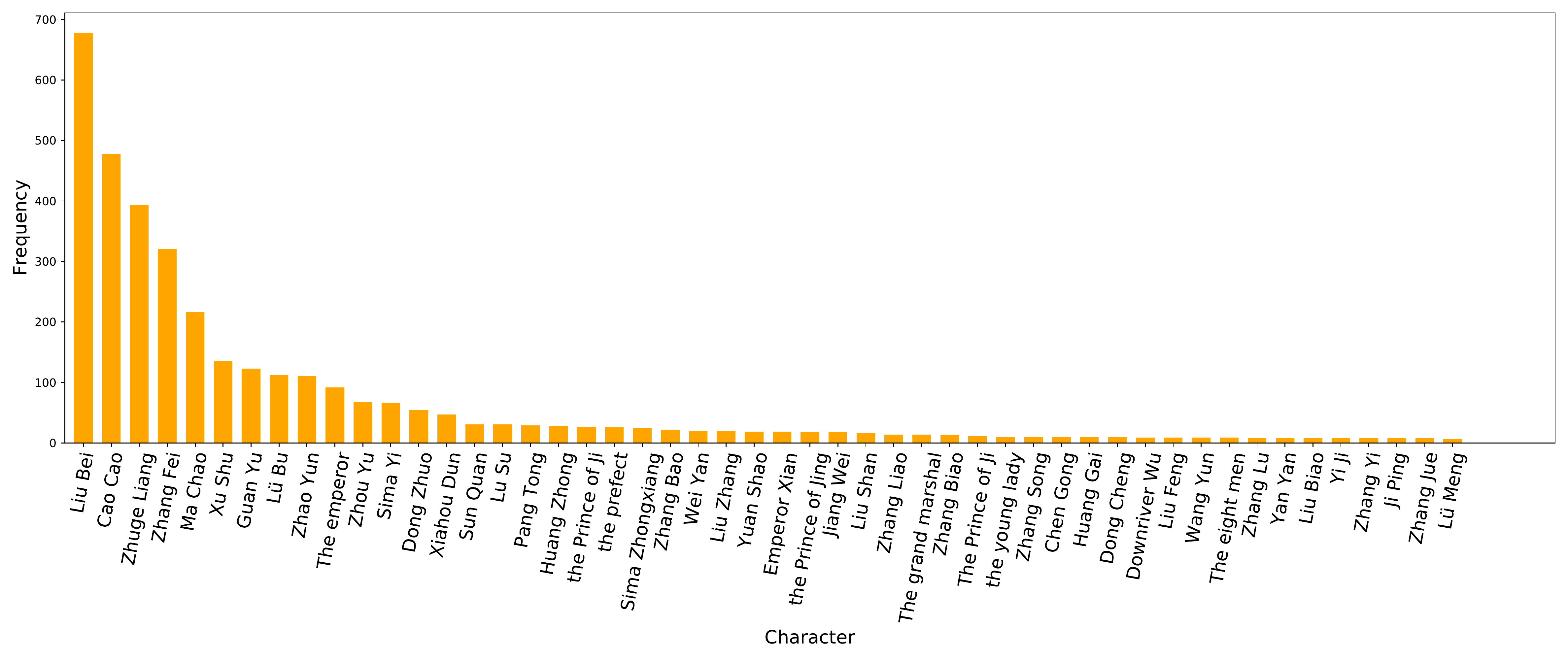}}
    \caption{The frequency of characters occurences.}
    \label{fig8}
\end{figure}

\begin{figure}
    \centering
    \includegraphics[scale=0.55]{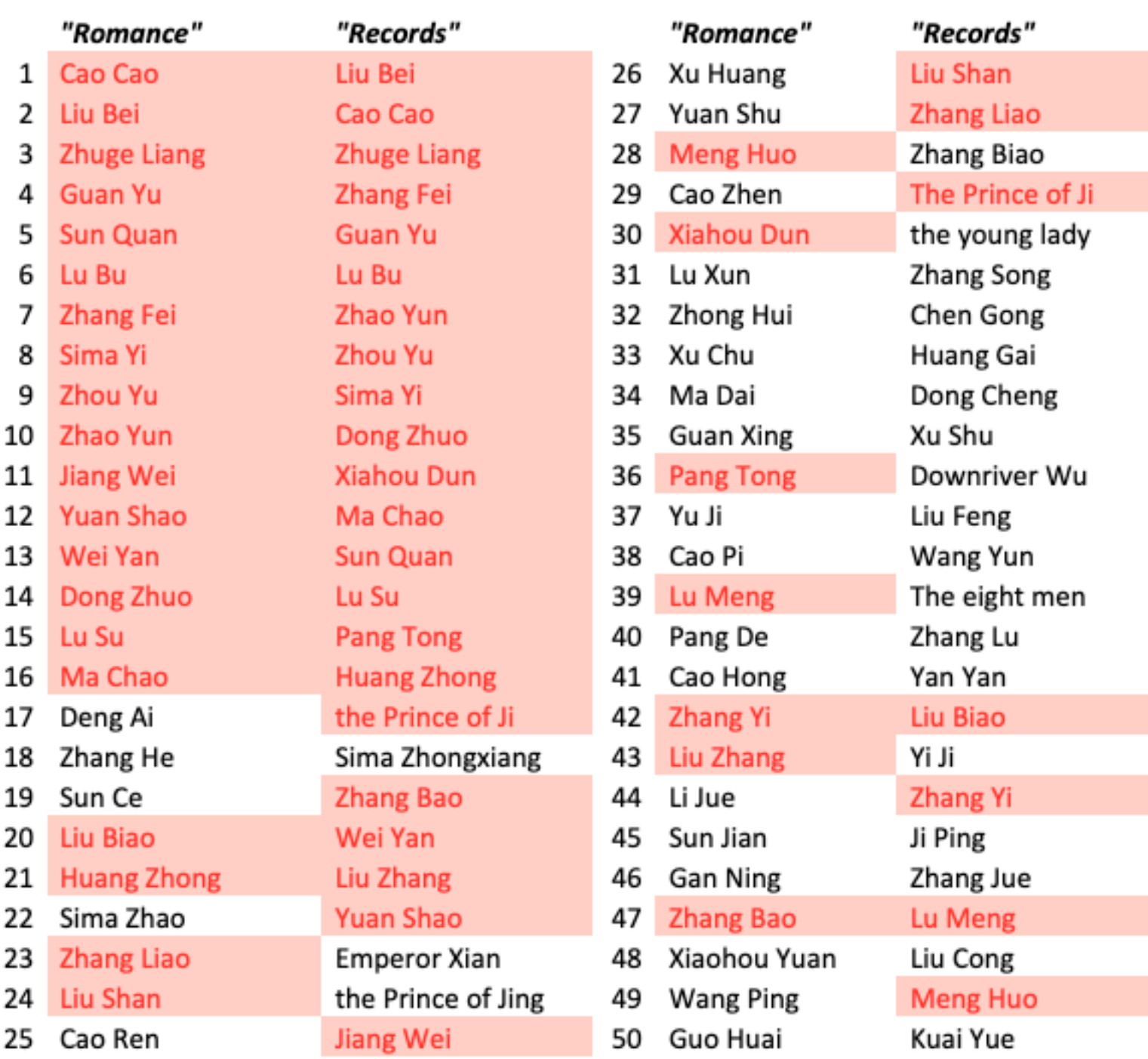}
    \caption{Similar characters that appear in both books’ top 50 ranks are highlighted in red.}
    \label{fig9}
\end{figure}

The results of the comparison is shown in Table \ref{tab4}, where the number of nodes of \textit{Romance} is four times of that of \textit{Records}. Additionally, the diameter of \textit{Romance} is 9 while the one of \textit{Records} is only 3. Consequently, there are more interactions (i.e., edges) in \textit{Romance}. It implies that the social network of Romance is grander than that of Records.

\begin{table}[htbp]
  \centering
  \caption{The number of nodes and edges, and the diameter of the two networks.}
    \begin{tabular}{lrrr}
    \toprule
          & \multicolumn{1}{l}{$\mathcal{N}$} & \multicolumn{1}{l}{$\mathcal{E}$} & \multicolumn{1}{l}{$d$} \\ \hline
    \textit{Romance} & 510   & 13,348 & 9 \\
    \textit{Records} & 128   & 8,986 & 3 \\
    \bottomrule
    \end{tabular}%
  \label{tab4}%
\end{table}%

\subsubsection{Similar Casts?}
Since the two books narrate the same story, we assume the casts of them (appearing characters) are similar. According to the statistical results, we find that while the number of characters involved in the \textit{Romance} is far larger than that of \textit{Records}, the former covers approximately 71.4\% characters of the latter. It indicates the similarity of casts between the two books is very high. In addition, we rank the top 50 most frequently appearing characters in the two books (See Figure \ref{fig8}) and find a 50\% coincidence referring to Figure \ref{fig9}. The protagonists (i.e., the top 20 characters) are notably similar; the top 3 most frequently appearing characters in both books are Liu Bei, Cao Cao,  and Zhuge Liang.

\subsubsection{Complex Network Features of the Story}
Previous studies indicate that novelistic literature usually involves a social network that is more complex and thus the literariness and the dramaticism can be greatly enriched \cite{waumans2015topology,elson2010extracting}. In this paper, two main topological features of the complex networks are considered, and related investigations are conducted on the two extracted social networks.

\paragraph{Small-world} Small-world is a complex network feature that describes a random network with a highly clustered structure. In a small-world network, most nodes are not neighbors of each other, yet the neighbors of some random nodes are probably going to be neighbors of one another, and most nodes can be reached from each other node by few jumps or steps. We can find out more homogeneity in the social structure when its social network possesses such a small-world feature. To measure the small-world feature in our extracted social networks, we introduce the two key metrics, namely, average clustering coefficient and average path length. Small-world networks are usually recognized as having large average path value length and low average clustering coefficient value. Moreover, an advanced metric, Small-World Index ($\mathrm{SWI}$) \cite{neal2017small}, is introduced. $\mathrm{SWI}$ is capable of quantifying the small-world feature, which can provide a more straightforward recognition. The calculation of $\mathrm{SWI}$ is
\begin{equation}
    \mathrm{SWI}=\frac{(L-{{L}_{l}})(C-{{C}_{r}})}{({{L}_{r}}-{{L}_{l}})({{C}_{l}}-{{C}_{r}})},
\end{equation}
where $C$ and $L$ are the clustering coefficient and average path length respectively, which are derived from the observed network (note that we compute them by Gephi in this work); $C_l$ and $L_l$ refer to the clustering coefficient and mean path length in a lattice reference network characterized by a high $C$ and $L$; Similarly, $C_r$ and $L_r$ refer to the clustering coefficient and mean path length in a random reference graph characterized by a low $C$ and $L$.

\begin{table}[htbp]
  \centering
  \caption{Average clustering coefficient, average path length, and small-world index of the two networks.}
    \begin{tabular}{lccc}
    \toprule
          & \multicolumn{1}{l}{Avg. Clustering coefficient} & \multicolumn{1}{l}{Avg. Path length} & \multicolumn{1}{l}{$\mathrm{SWI}$} \\ \hline
    Romance & 0.326 & 2.977 & 0.8624 \\
    Records & 0.647 & 2.127 & 1.56247 \\
    \bottomrule
    \end{tabular}%
  \label{tab5}%
\end{table}%

From the results shown in Table \ref{tab5}, we can observe that the \textit{Records} has a significantly higher average path value and smaller average clustering coefficient compared to those of \textit{Romance}. Especially, the results of the calculated $\mathrm{SWI}$ indicate that the $\mathrm{SWI}$ of Records (1.562) is higher than that of Romance (0.862), thereby quantifiably confirming our assumption. Literature that focuses on a single character or a group of characters presents a higher $\mathrm{SWI}$ than those focused on a mass of characters. \textit{Romance} focuses on a few protagonists, features a much higher $\mathrm{SWI}$ than the \textit{Records}, where the story follows several protagonists. It implies that \textit{Romance} focuses more on storytelling around several characters rather than epic depiction.

\begin{figure}
    \centering
    \subfigure[Degree distribution of the network in \textit{Romance}.]{
    \includegraphics[scale=0.3]{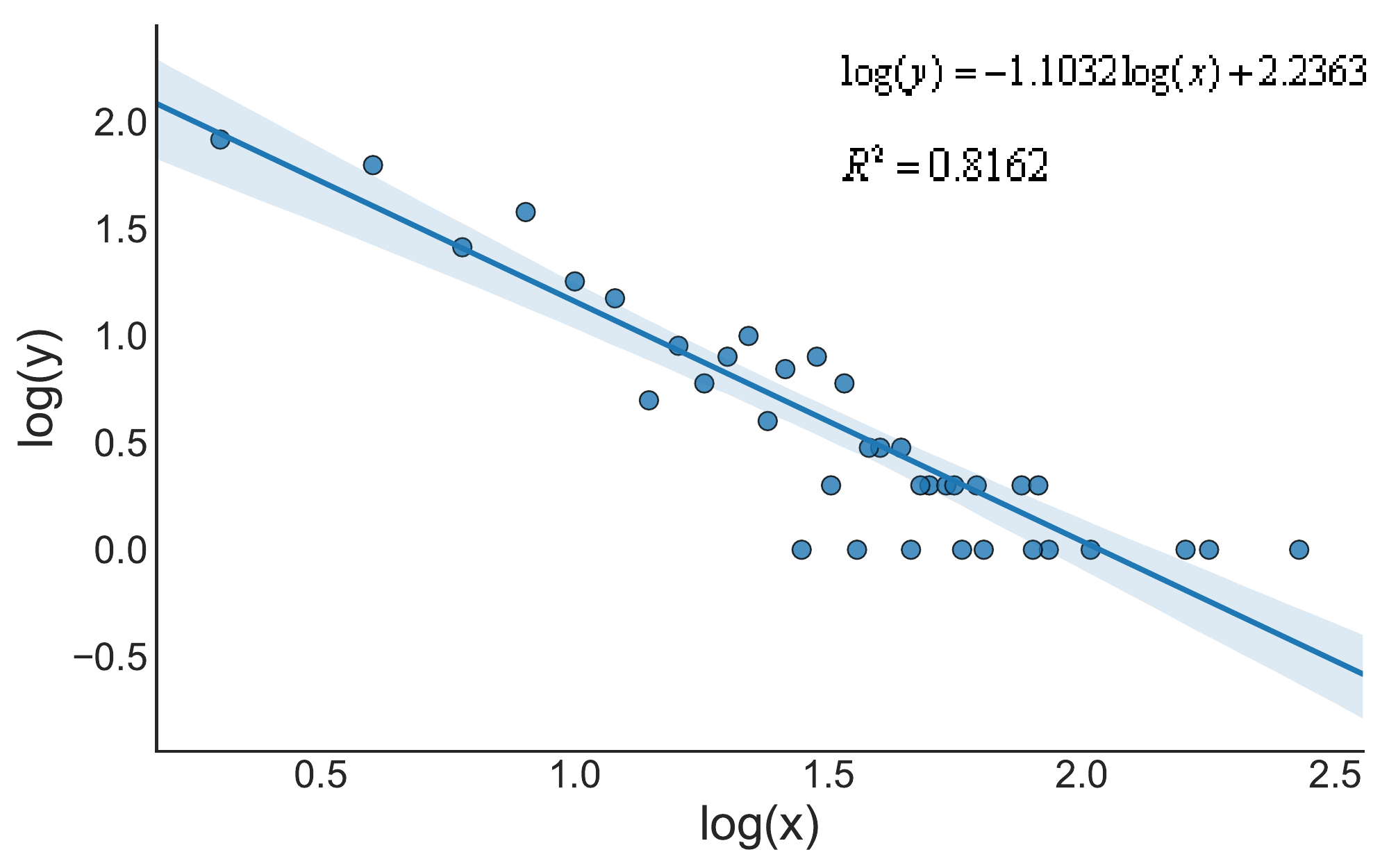}}
    \quad
    \subfigure[Degree distribution of the network in \textit{Records}.]{
    \includegraphics[scale=0.3]{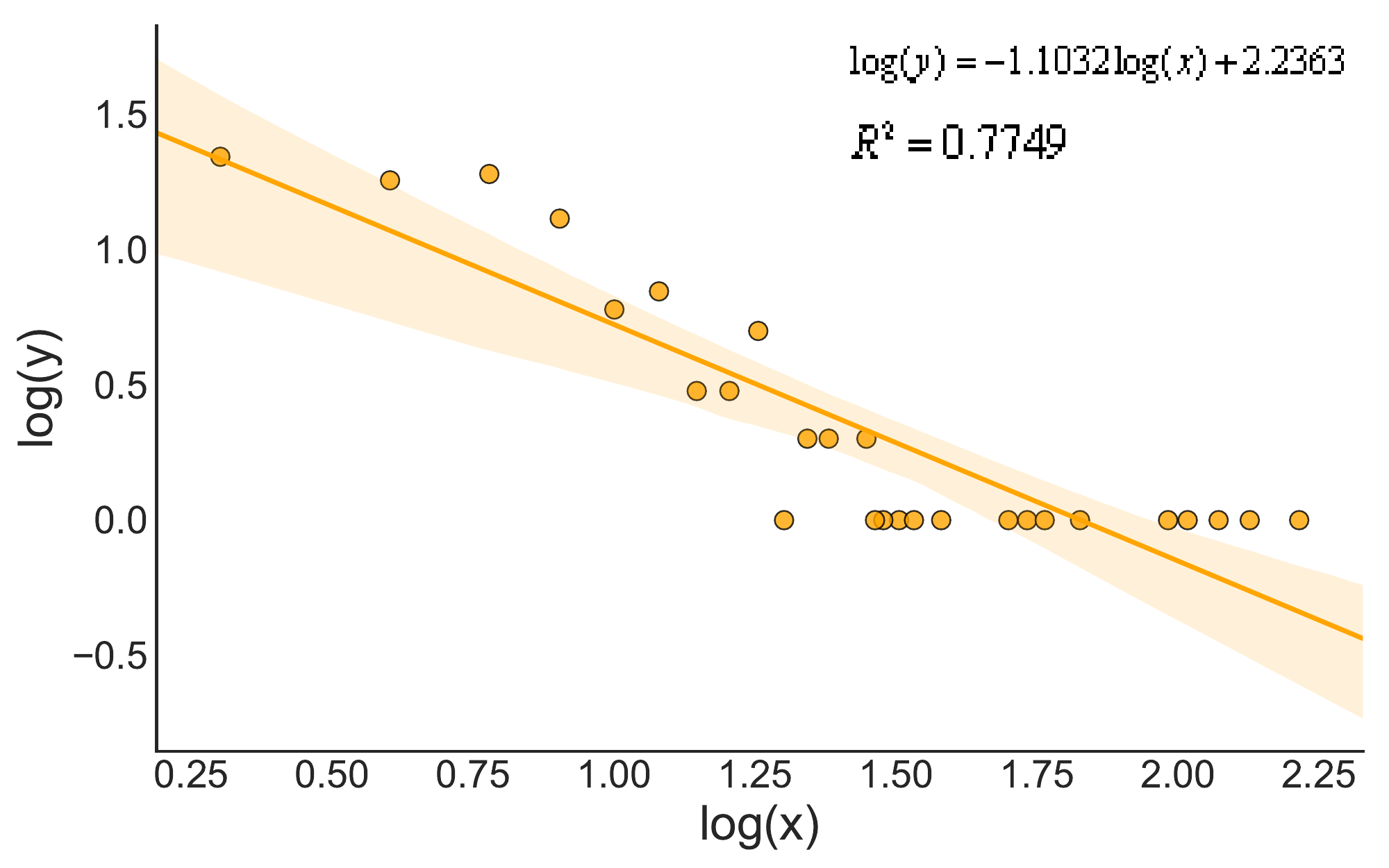}}
    \caption{Degree distribution.}
    \label{fig10}
\end{figure}

\paragraph{Scale-free} Scale-free describes a network whose degree distribution follows a power law. It reveals the Pareto principle that 20\% of individuals commonly hold 80\% of the total resources in a society, a.k.a., ``the rich get richer'' \cite{chakraborti2009variational}. To investigate the scale-free feature of the two networks, we demonstrate the degree distribution of nodes in them, as shown in Figure \ref{fig10}, where $x$ is the degree of a node and $y$ is the number of nodes which possess this degree. 

From the results, we can observe a salient power-law distribution in the diagrams of both networks. The satisfaction of the power-law indicates that networks of the two books are both scale-free. However, the distribution of \textit{Romance} has a more significant coefficient of determination ($R^2$: 0.8162 > 0.7749) than that of Records, which means that Romance is relatively more in line with this law.

\paragraph{Rich-club Coefficient} A number of scale-free networks exhibit a “rich-club” feature, indicating that a small number of nodes possessing a large number of edges also connect well to one another \cite{zhou2004rich,alstott2014unifying}. The Rich-Club Coefficient is used to measure this feature, which can be computed by
\begin{equation}
    \phi(k)=\frac{2{\mathcal{E}_{>k}}}{{\mathcal{N}_{>k}}(\mathcal{N}_{>k}-1)}
\end{equation}
where $\mathcal{N}_{>k}$ is the number of nodes whose degree is not less than $k$, and $\mathcal{E}_{>k}$ is the actual number of edges among the nodes whose degree are not less than $k$; $\phi(k)$ is the ratio between the number of edges that exist among the nodes that have a degree larger than $k$ and the total possible number among them. Considering the different sizes of the two networks, we compare the ratio of nodes that can form a fully connected network ($\phi(k)$ = 100\%) deduced by the cut-off degree observed, which can be calculated by
\begin{equation}
    {r_{fc}} = \frac{{{k_{\phi (k) = 100\% }}}}{\mathcal{N}},
\end{equation}
where $k_{\phi (k) = 100\% }$ is the minimum $k$ which make $\phi (k) = 100\%$, and $N$ is the number of nodes in the network.

The calculated $r_{fc}$ are 5.09\% (\textit{Romance}) and 20.31\% (\textit{Records}), respectively. It reveals that the top 5\% rich nodes in \textit{Romance} can approximately form a fully connected network, whereas the number has to be approximately the top 20\% in \textit{Records}. It can be concluded that both networks have a rich-club feature, which is more significant in \textit{Romance}. These results reveal that despite more characters appearing in \textit{Romance} than in \textit{Records}, the story always revolves around a few protagonists in \textit{Romance}.

\begin{figure}
    \centering
    \includegraphics[scale=0.19]{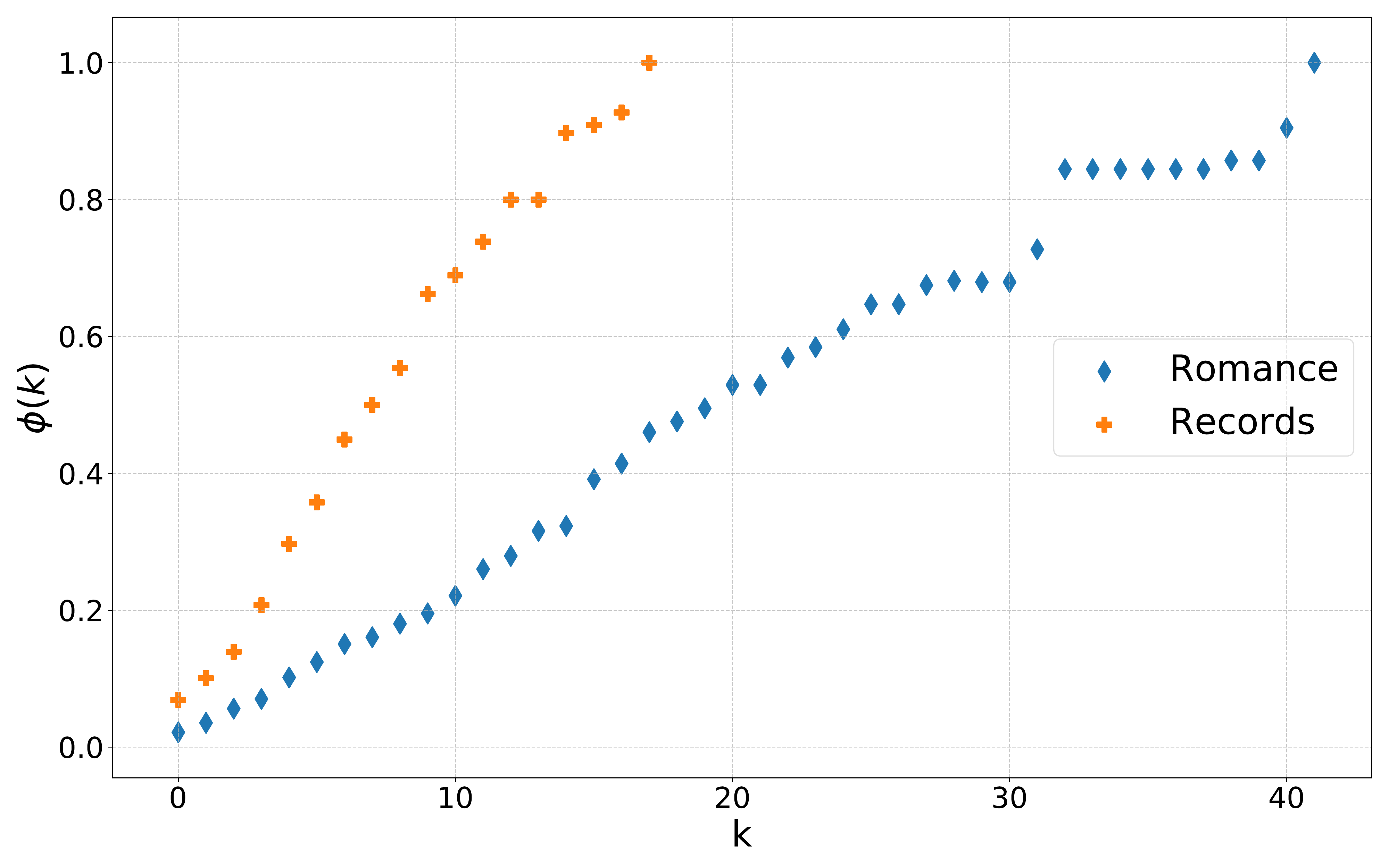}
    \caption{Rich-club coefficients of the two networks.}
    \label{fig11}
\end{figure}

\subsubsection{History or Romance: Which is More Dramatic?}
\begin{figure}
    \centering
    \subfigure[Number of nodes.]{
    \includegraphics[scale=0.16]{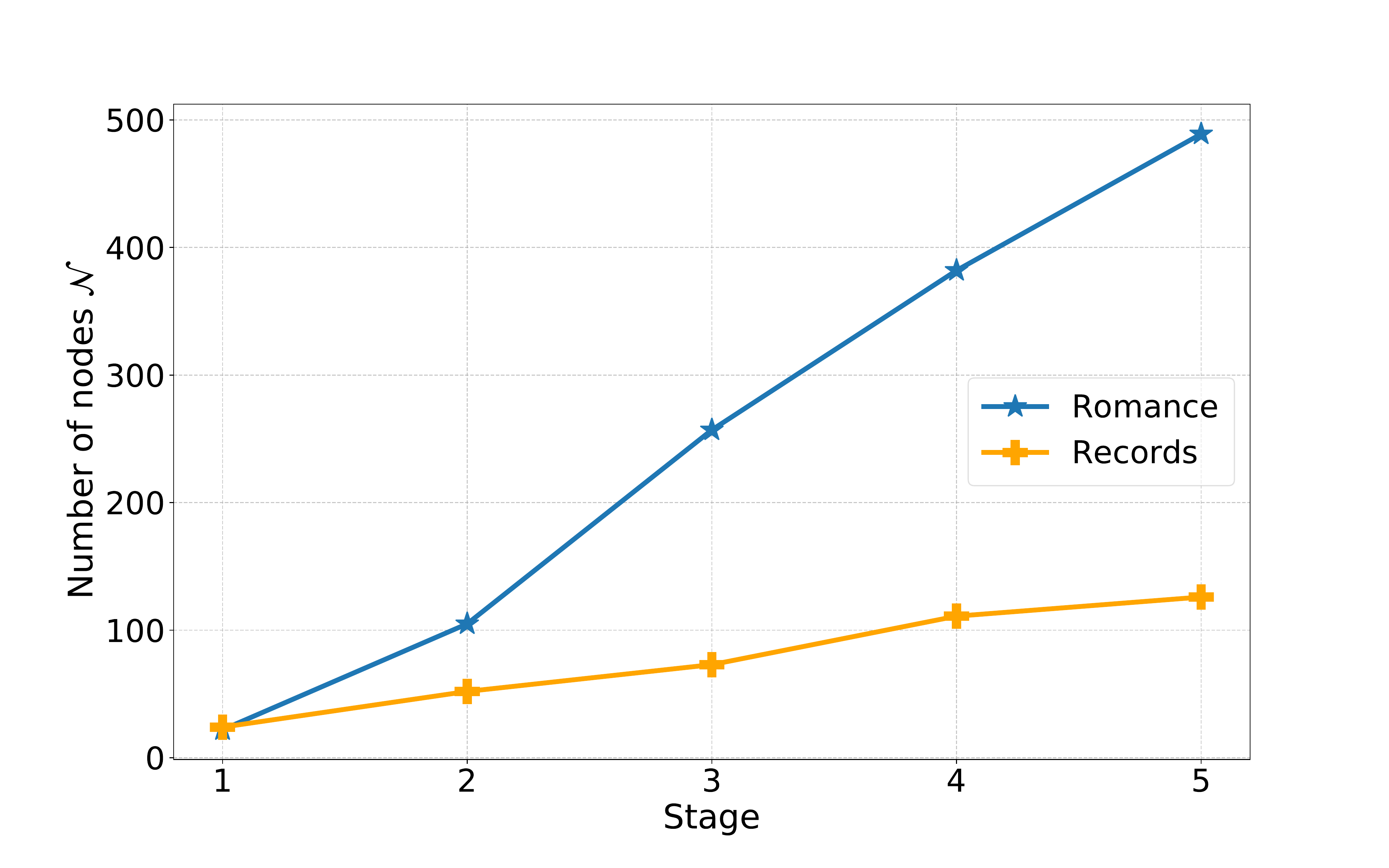}}
    \quad
    \subfigure[Average degree.]{
    \includegraphics[scale=0.16]{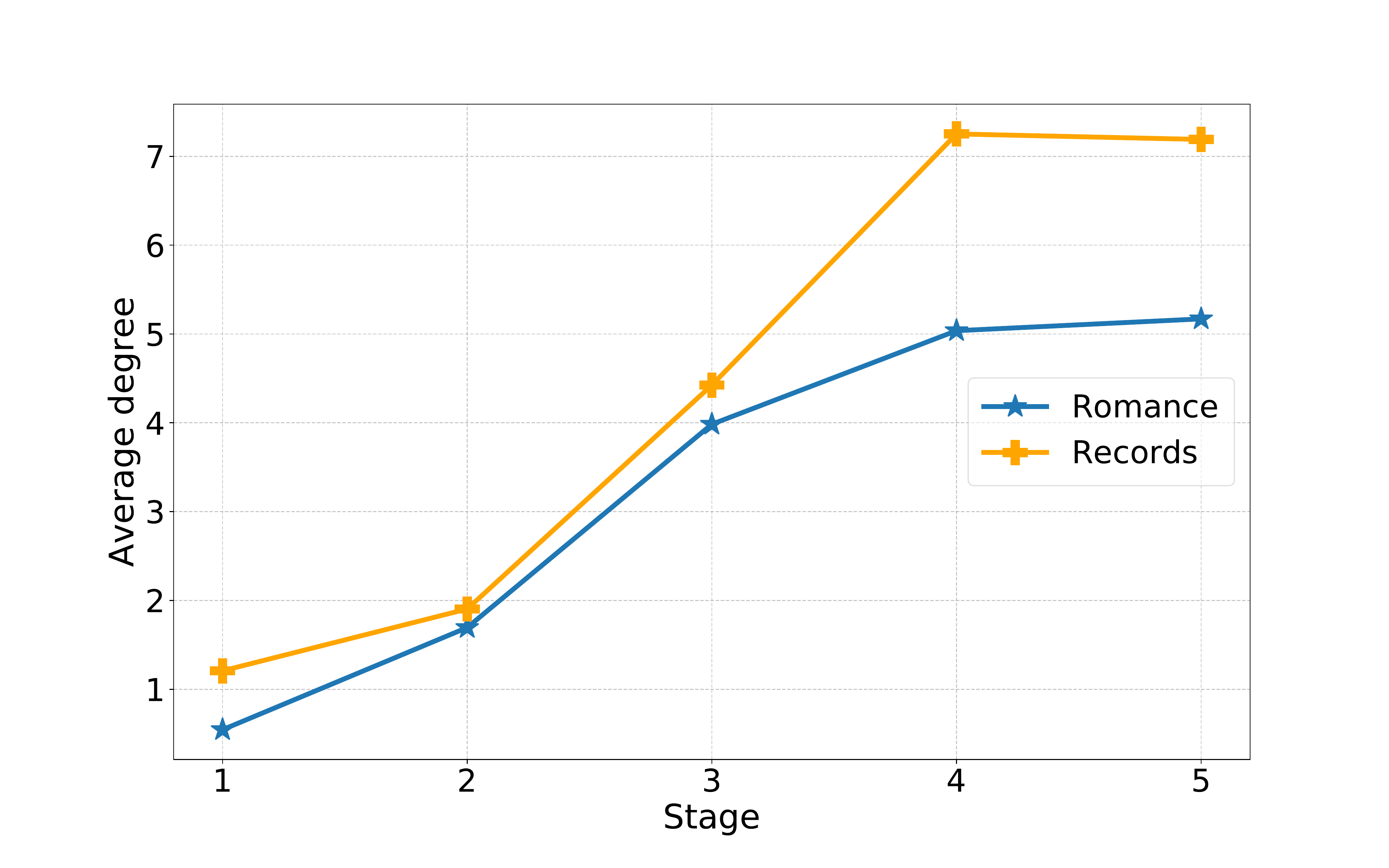}}
    \quad
    \subfigure[Average path length.]{
    \includegraphics[scale=0.16]{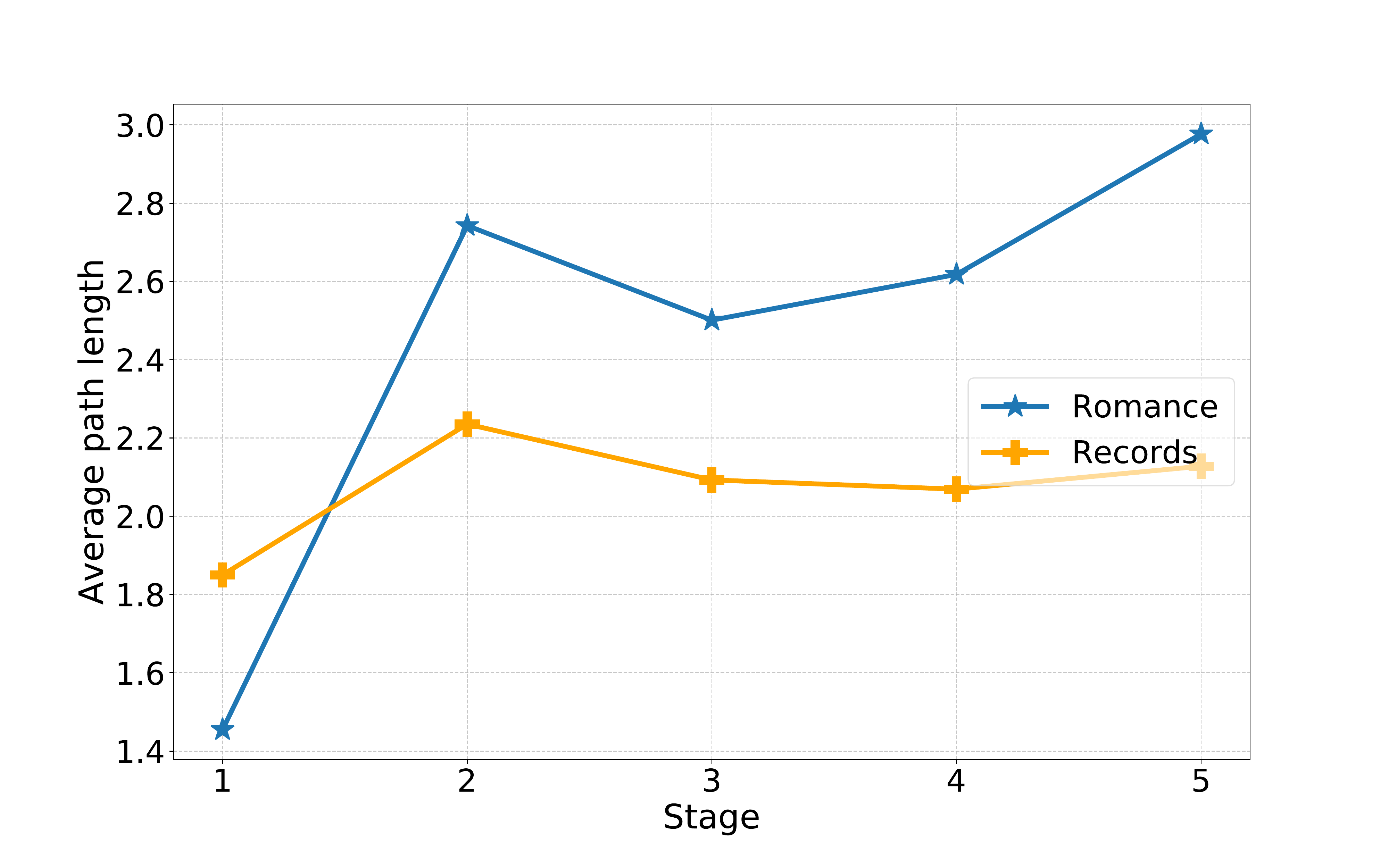}}
    \quad
    \subfigure[Density.]{
    \includegraphics[scale=0.16]{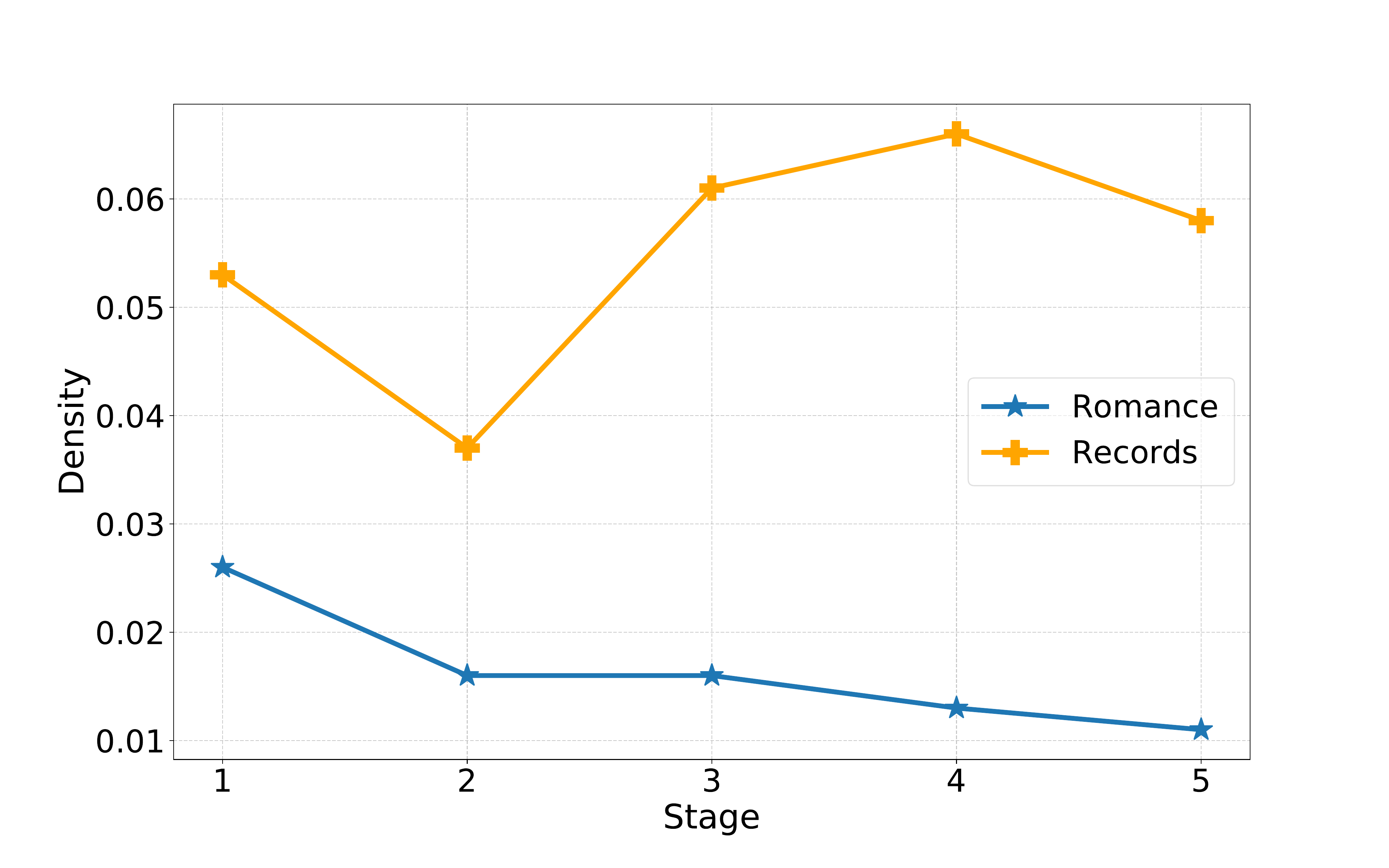}}
    \quad
    \subfigure[Average clustering coefficient.]{
    \includegraphics[scale=0.16]{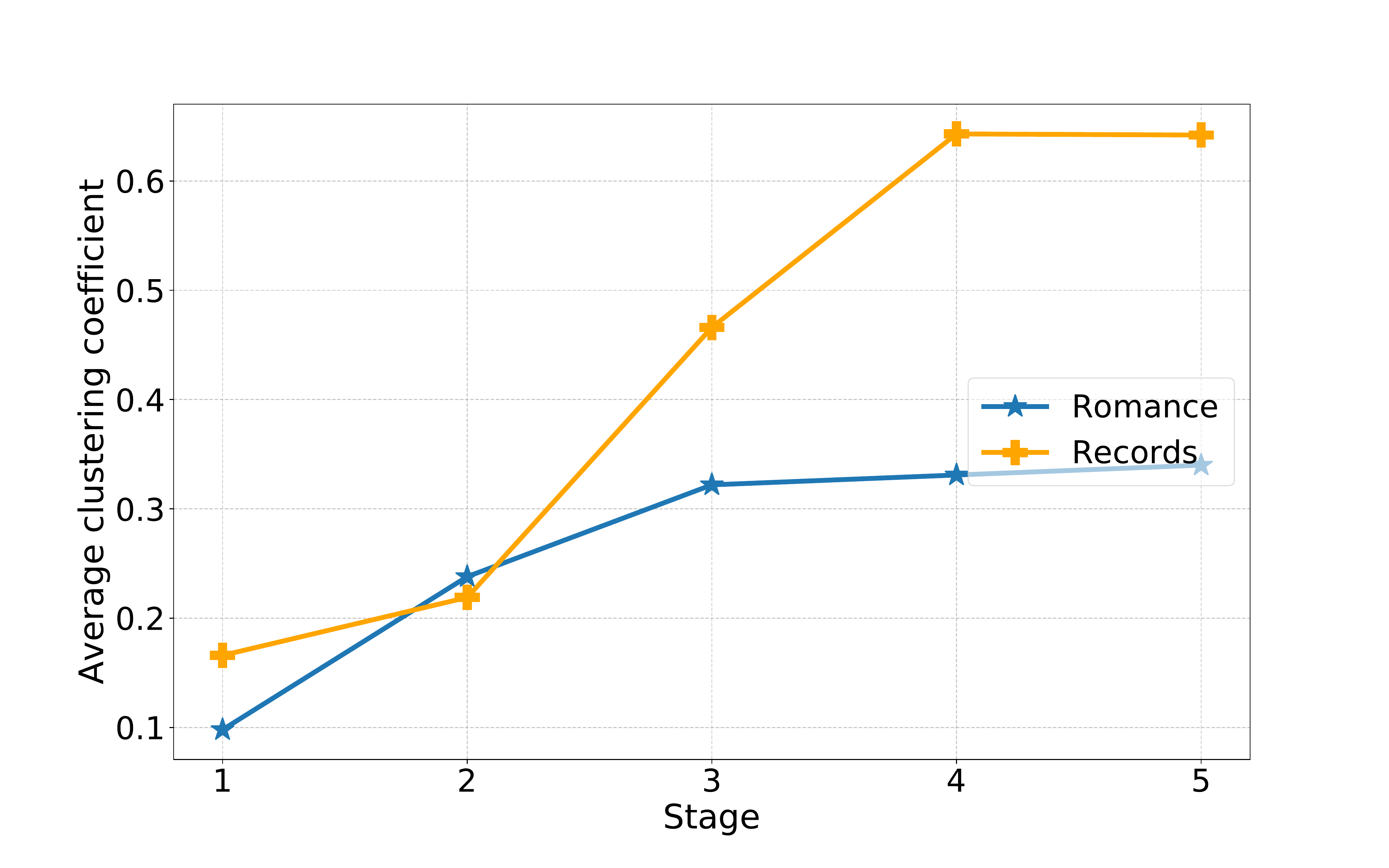}}
    \caption{Dynamic growth of the two networks (5 stages).}
    \label{fig12}
\end{figure}

Dramatic changes make stories splendid. The rise and fall of warlords constantly change the social structure of the story of the Three Kingdoms across all stages. To study the growth of the social structure, we investigate the social network according to the idea introduced in Section \ref{method_dynamic}. As shown in Figure \ref{fig12}, five metrics are adopted to observe the dynamic change of the networks.

Interesting phenomena are found in the results as below. The average node growth rate in \textit{Romance} and \textit{Records} are 147\% and 63\%, respectively. This suggests that \textit{Romance} has dramatic changes in terms of the number of characters, and the appearance of characters on each stage is overwhelming. The density comparison indicates that \textit{Romance} has a larger network size through all the stages, yet its density is lower, where the gap in the last three stages is especially notable. The change of the average degree of two networks follows a similar pattern, demonstrating that they both increase at the beginning and then reach a plateau. \textit{Records} has a considerably larger average path length and lower average clustering coefficient in the majority of five stages except in the first two, which match its better performance in small-worldliness. Overall, we can observe that the growth of the social network in \textit{Romance} is more rapid. Comparatively, Records has a tight, and gradually clustered network.

\subsection{Network Feature on Specific Characters}
In this subsection, we assess the network feature on specific characters. While the story of the Three Kingdoms involves numerous forces, the main focus is the three force blocs, i.e., Wei, Shu, and Wu. Therefore, their respective sovereigns, namely, Cao Cao (Wei), Liu Bei (Shu), and Sun Quan (Wu), are chosen as the targets for our character-centric investigation.

\subsubsection{Who is the most influential?}
In the story of the Three Kingdoms, the personal influence of each sovereign considerably represents the influence of the forces they possess. Given this kind of influence in a social network, the sovereigns’ interactions with other characters reflect their influence. Three related metrics are introduced to compare their influence:

\begin{itemize}
    \item \textbf{Degree:} Degree or degree centrality is a basic measure that counts the number of neighbors that a node (character) has. The weighted degree is additionally considered, which is calculated by considering the number of interactions that occur between two characters.
    
    \item \textbf{Closeness centrality:} Closeness centrality measures the extent of closeness of a node to a network. It is calculated as the reciprocal of the sum of the length of the shortest paths between the node and all other nodes in the graph. Its formula is expressed as
    \begin{equation}
    C(i)=\frac{1}{\sum\nolimits_{j}{d(i,j)}},  
    \end{equation}
    where $C(i)$ is the closeness centrality of node $i$, and $d(i,j)$ denotes the distance between node $i$ and node $j$.
    
    \item \textbf{Betweenness centrality:} For each pair of nodes in a network, at least one shortest path exists between nodes, wherein either the number of edges that the path passes through (for unweighted networks) or the sum of the weights of the edges (for weighted networks) is minimized. Betweenness centrality is a measure of the number of the shortest path that passes through a node. Denoted by $g(v)$, the betweenness centrality of node $v$ can be calculated by
    \begin{equation}
    g(v)=\sum\nolimits_{i\ne v\ne j}{\frac{{{\sigma }_{ij}}(v)}{{{\sigma }_{ij}}}},
    \end{equation}
    where $\sigma_{ij}$ is total number of shortest paths from node $i$ to node $j$, and $\sigma_{ij}(v)$ is the number of those paths that pass through node $v$. A character’s property of ``bridge'' can be measured by betweenness centrality.
\end{itemize}

Table \ref{tab6} presents the results of \textit{Romance} and Table \ref{tab7} shows the results of \textit{Records}. In \textit{Romance}, Cao Cao exhibits the highest measures of the four metrics, followed by Liu Bei, whereas Sun Quan has the lowest measures. In \textit{Records}, Liu Bei leads the performance instead of Cao Cao, and Sun Quan is far behind them. This can support us to conclude that Cao Cao is the most influential of the three sovereigns in \textit{Romance}, and Liu Bei is the one in \textit{Records}, and the influence of Sun Quan is lower than the other two lords in both books.

\begin{table}[htbp]
  \centering
  \caption{The degree and centrality of Cao Cao, Liu Bei, and Sun Quan in the network of \textit{Romance}.}
    \begin{tabular}{lrrrr}
    \toprule
          & \multicolumn{1}{l}{Degree} & \multicolumn{1}{l}{Weighted degree} & \multicolumn{1}{l}{Closeness centrality} & \multicolumn{1}{l}{Betweenness centrality} \\ \hline
    Cao Cao & 268   & 3,442 & 0.565371 & 18,765.21 \\
    Liu Bei & 178   & 2,270 & 0.506329 & 7,891.51 \\
    Sun Quan & 80    & 908   & 0.448808 & 1,688.14 \\
    \bottomrule
    \end{tabular}%
  \label{tab6}%
\end{table}%

\begin{table}[htbp]
  \centering
  \caption{The degree and centrality of Cao Cao, Liu Bei, and Sun Quan in the network of \textit{Records}.}
    \begin{tabular}{lrrrr}
    \toprule
          & \multicolumn{1}{l}{Degree} & \multicolumn{1}{l}{Weighted degree} & \multicolumn{1}{l}{Closeness centrality} & \multicolumn{1}{l}{Betweenness centrality} \\ \hline
    Liu Bei & 164   & 2,753 & 0.778523 & 4,731.94 \\
    Cao Cao & 134   & 1,920 & 0.707317 & 2,764.84 \\
    Sun Quan & 24    & 326   & 0.527273 & 20.007937 \\
    \bottomrule
    \end{tabular}%
  \label{tab7}%
\end{table}%

\subsection{Sentiment Analysis on Characters}
\begin{figure}
    \centering
    \subfigure[Cao Cao.]{
    \includegraphics[scale=0.08]{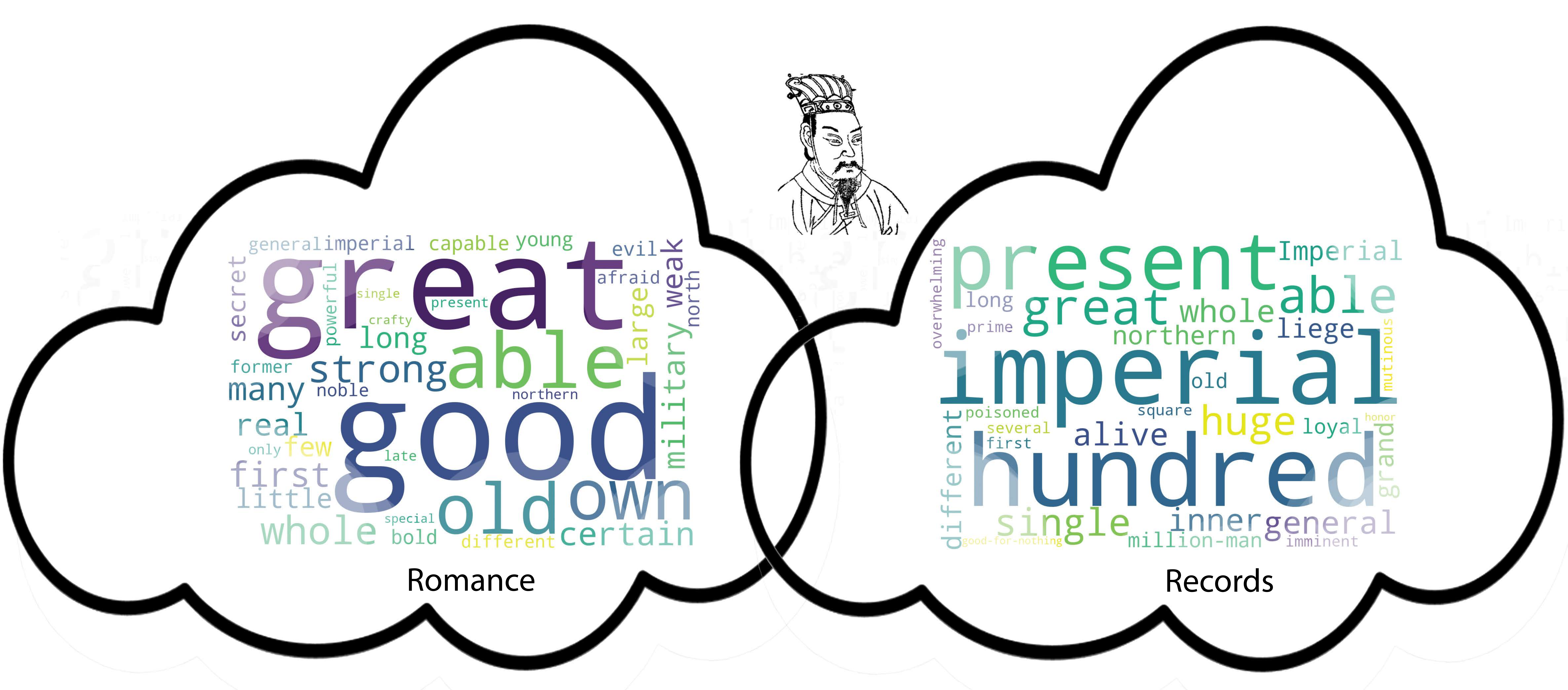}}
    \quad
    \subfigure[Liu Bei.]{
    \includegraphics[scale=0.08]{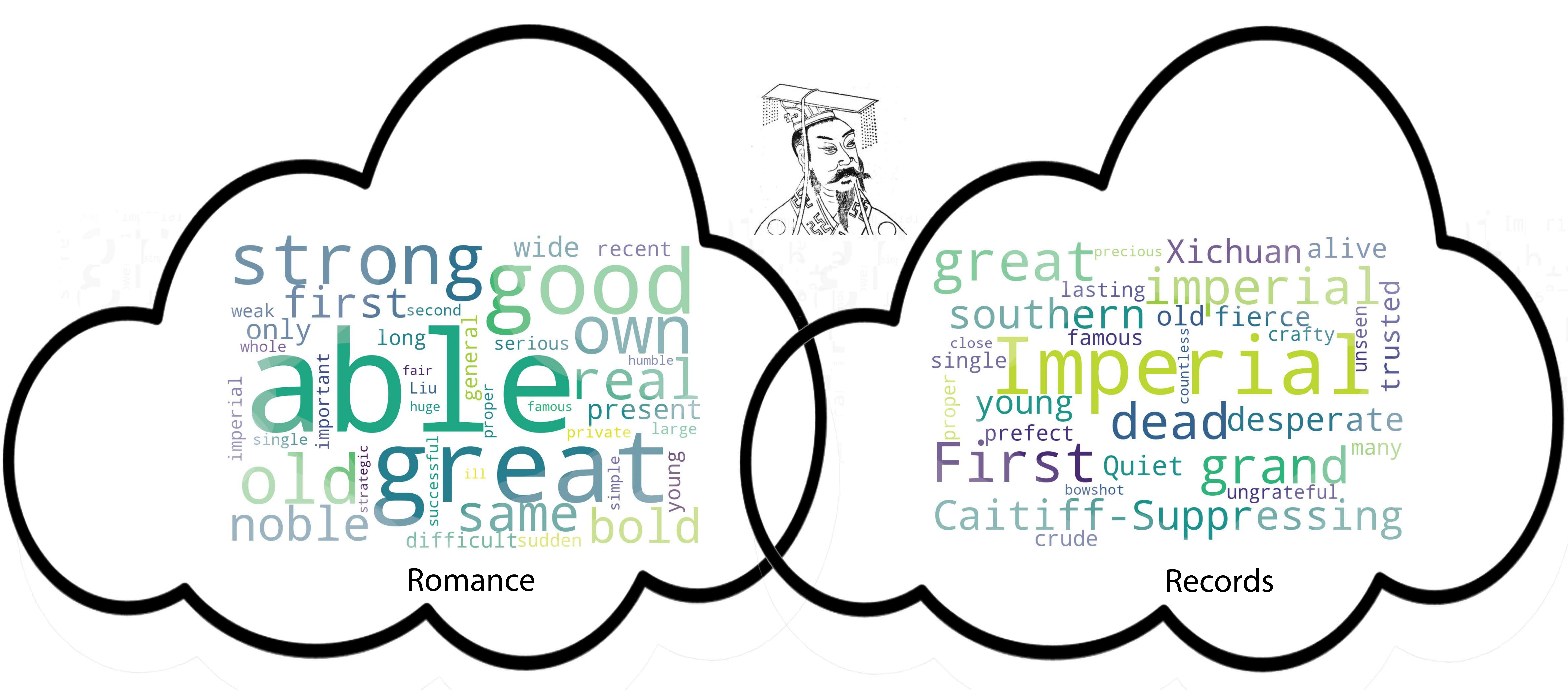}}
    \caption{Word clouds of the evaluative words on Cao Cao and Liu Bei.}
    \label{fig13}
\end{figure}
Generally, historical records tend to lean toward objectivity, whereas fictional novels contain subjective emotions. The creation of Romance began approximately toward the end of the Yuan Dynasty, which was a dark era for common people. The dissatisfaction with the ruling class can be reflected by the impressionable attitude of people to some forces (e.g., Shu) in the story of the Three Kingdoms. In this context, the author of \textit{Romance} emotionally depicted a series of characters who are different in actual history. This phenomenon substantially occurs to Liu Bei and Cao Cao, which are lords of Shu and Wei, respectively. Our investigation focuses on these two characters from the point of their evaluating words.

\subsubsection{A ``Taste'' of the Character Sentiment}
\label{taste}
We commence by collecting and ranking the evaluative words to Liu Bei and Cao Cao. Specifically, we present the results by adopting the word cloud, as shown in Figure \ref{fig13}. As the word cloud shown in Figure \ref{fig13}, A sketchy sentimental opinion on Cao Cao and Liu Bei can be obtained. For example, in \textit{Romance}, evaluative words such as “great” and “able” are mentioned for both two lords. However, we in addition find words such as “crafty” and “evil” on Cao Cao and “humble” on Liu Bei, which reveals the difference. Moreover, more negative words are obviously found about Cao Cao in \textit{Romance} than in \textit{Records}. While this observation cannot bring us to the conclusion that the authors of the two books have an evident preference to a character, we can at least find the there exists differences regarding the depiction of the same character in the two books.

\begin{figure}
    \centering
    \includegraphics[scale=0.19]{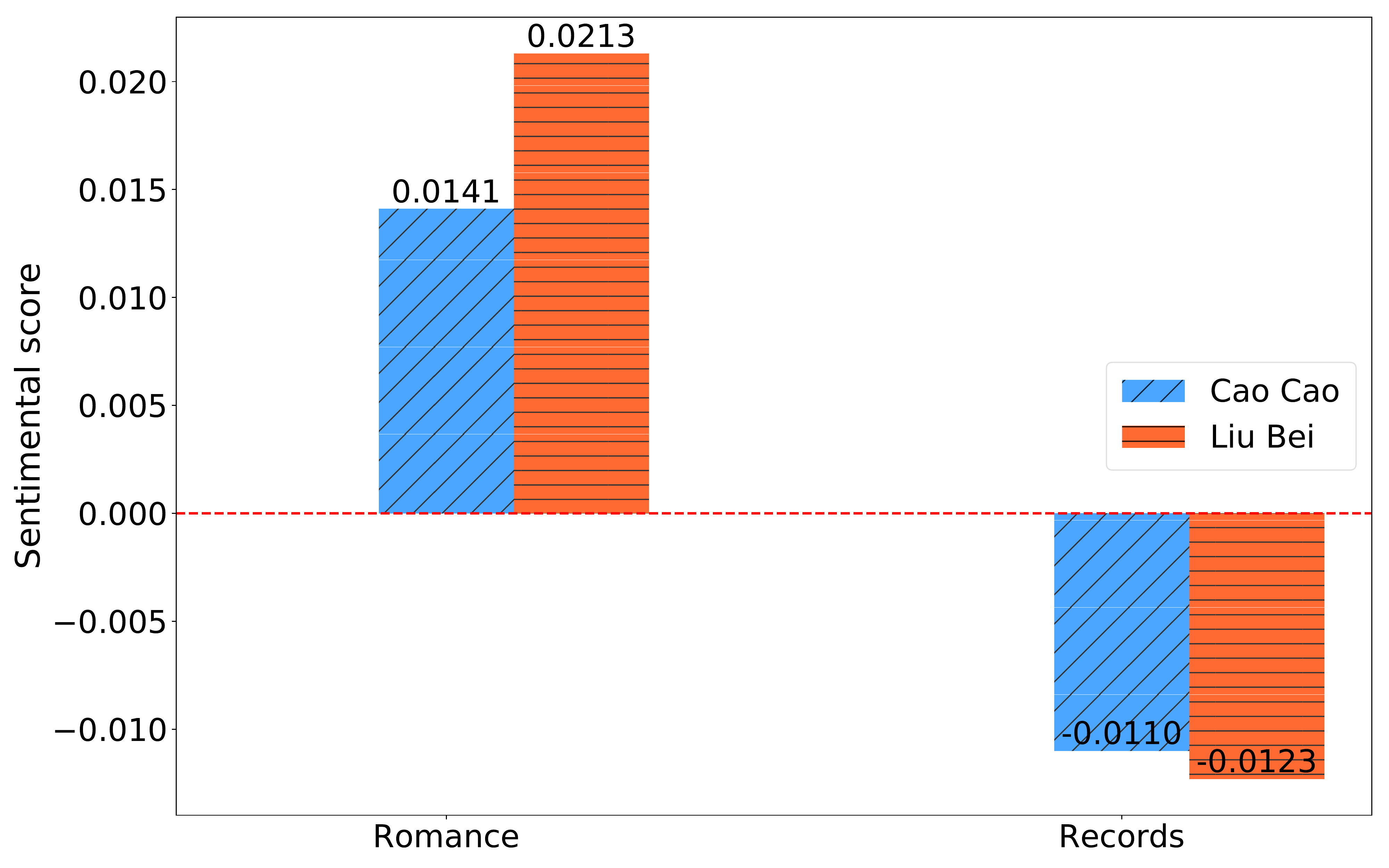}
    \caption{SentiWordNet scores of Cao Cao and Liu Bei.}
    \label{fig14}
\end{figure}

\subsubsection{Sentimental Quantification}
For a better understanding of the evaluative words to , we conduct a quantitative comparison . Particularly, we introduce a sentimental scoring metric, SentiWordNet \cite{baccianella2010sentiwordnet}. SentiWordNet score can be calculated by subtracting both polarities (positive and negative) of each token and subsequently calculating them:
\begin{equation}
    score=\frac{\sum\nolimits_{i=1}^{n}{(posScore_i-negScore_i)}}{n}
\end{equation}
where $n$ denotes the number of involved evaluative words, $posScore_i$ and $negScore_i$ are the positive and negative scores of word $i$ provided by SentiWordNet. The criterion of SentiWordNet gives Negative (i.e., ˗1), Neutral (i.e., 0) and Positive (i.e., 1) for users to classify the word.

Figure \ref{fig14} implies that Cao Cao’s score is lower than that of Liu Bei in \textit{Romance}. Nonetheless, the score of Cao Cao is higher in \textit{Records}. This finding is consistent with the subjective perception obtained in Section \ref{taste}. In addition, the scores of the two characters are both higher in \textit{Romance} than in \textit{Records}. This possibly reveals the different sentimental tones of the authors’ wording in the narrative.

\section{Conclusion}
\label{sec:conclusion}
Surrounding on the Story of Three Kingdoms, this paper revives the research on digital humanities, which seeks to digitize working procedures of sociologists and historians in the field of humanities by using state-of-the-art data science technologies. 

An algorithm is developed to extract social networks of stories narrated in two books w.r.t. the Three Kingdoms. Particularly, the advanced NLP model BERT is employed in our character identification work, and a satisfying outcome is obtained. Subsequently, we conduct a series of topological analysis to quantify and characterize the extracted social networks, where we additionally present a quantitative comparison between the two books. Specifically, network topological features, such as small-world, scale-free, and centrality of specific characters, are measured. The results reveal that the social network is more entangled in the narrative of the \textit{Romance} than that of the \textit{Records}, especially, more protagonist-oriented. Moreover, this provides a quantitative reference for the macro (e.g., structural features of a story) and micro levels (e.g., the influence or sentiment of a specific character), and the extent of the grandness vividness of a story can be expressed scientifically.

This work can help both researchers and non-expert readers gain an insight into the story of the Three Kingdoms and the procedure of its digital analysis. Moreover, numerous involved sub-works can be refined in the future. First, the definition of interactions between characters is coarse-grained. Second, a mere five-slice dynamic network is built in this project, and hopefully, a large-scale dynamic network, which can incorporate hundreds even thousands of slices, can be obtained if the story is subdivided in fine granularity, for instance, year-to-year or day-to-day.

\section{Acknowledgments}

% Identification of funding sources and other support, and thanks to
% individuals and groups that assisted in the research and the
% preparation of the work should be included in an acknowledgment
% section, which is placed just before the reference section in your
% document.

% \begin{acks}
% \end{acks}

\bibliographystyle{ACM-Reference-Format}
\bibliography{sample-base}


\begin{thebibliography}{42}

%%% the Web address.

\ifx \showCODEN    \undefined \def \showCODEN     #1{\unskip}     \fi
\ifx \showDOI      \undefined \def \showDOI       #1{#1}\fi
\ifx \showISBNx    \undefined \def \showISBNx     #1{\unskip}     \fi
\ifx \showISBNxiii \undefined \def \showISBNxiii  #1{\unskip}     \fi
\ifx \showISSN     \undefined \def \showISSN      #1{\unskip}     \fi
\ifx \showLCCN     \undefined \def \showLCCN      #1{\unskip}     \fi
\ifx \shownote     \undefined \def \shownote      #1{#1}          \fi
\ifx \showarticletitle \undefined \def \showarticletitle #1{#1}   \fi
\ifx \showURL      \undefined \def \showURL       {\relax}        \fi
% invisible to TeX
\providecommand\bibfield[2]{#2}
\providecommand\bibinfo[2]{#2}
\providecommand\natexlab[1]{#1}
\providecommand\showeprint[2][]{arXiv:#2}

\bibitem[\protect\citeauthoryear{Alstott, Panzarasa, Rubinov, Bullmore, and
  V{\'e}rtes}{Alstott et~al\mbox{.}}{2014}]%
        {alstott2014unifying}
\bibfield{author}{\bibinfo{person}{Jeff Alstott}, \bibinfo{person}{Pietro
  Panzarasa}, \bibinfo{person}{Mikail Rubinov}, \bibinfo{person}{Edward~T
  Bullmore}, {and} \bibinfo{person}{Petra~E V{\'e}rtes}.}
  \bibinfo{year}{2014}\natexlab{}.
\newblock \showarticletitle{A unifying framework for measuring weighted rich
  clubs}.
\newblock \bibinfo{journal}{\emph{Scientific reports}}  \bibinfo{volume}{4}
  (\bibinfo{year}{2014}), \bibinfo{pages}{7258}.
\newblock


\bibitem[\protect\citeauthoryear{Baccianella, Esuli, and
  Sebastiani}{Baccianella et~al\mbox{.}}{2010}]%
        {baccianella2010sentiwordnet}
\bibfield{author}{\bibinfo{person}{Stefano Baccianella},
  \bibinfo{person}{Andrea Esuli}, {and} \bibinfo{person}{Fabrizio Sebastiani}.}
  \bibinfo{year}{2010}\natexlab{}.
\newblock \showarticletitle{Sentiwordnet 3.0: an enhanced lexical resource for
  sentiment analysis and opinion mining.}. In \bibinfo{booktitle}{\emph{Lrec}},
  Vol.~\bibinfo{volume}{10}. \bibinfo{pages}{2200--2204}.
\newblock


\bibitem[\protect\citeauthoryear{Basole and Bellamy}{Basole and
  Bellamy}{2014}]%
        {basole2014visual}
\bibfield{author}{\bibinfo{person}{Rahul~C Basole} {and}
  \bibinfo{person}{Marcus~A Bellamy}.} \bibinfo{year}{2014}\natexlab{}.
\newblock \showarticletitle{Visual analysis of supply network risks: Insights
  from the electronics industry}.
\newblock \bibinfo{journal}{\emph{Decision Support Systems}}
  \bibinfo{volume}{67} (\bibinfo{year}{2014}), \bibinfo{pages}{109--120}.
\newblock


\bibitem[\protect\citeauthoryear{Bastian, Heymann, Jacomy,
  et~al\mbox{.}}{Bastian et~al\mbox{.}}{2009}]%
        {bastian2009gephi}
\bibfield{author}{\bibinfo{person}{Mathieu Bastian}, \bibinfo{person}{Sebastien
  Heymann}, \bibinfo{person}{Mathieu Jacomy}, {et~al\mbox{.}}}
  \bibinfo{year}{2009}\natexlab{}.
\newblock \showarticletitle{Gephi: an open source software for exploring and
  manipulating networks.}
\newblock \bibinfo{journal}{\emph{Icwsm}} \bibinfo{volume}{8},
  \bibinfo{number}{2009} (\bibinfo{year}{2009}), \bibinfo{pages}{361--362}.
\newblock


\bibitem[\protect\citeauthoryear{Bellamy and Basole}{Bellamy and
  Basole}{2013}]%
        {bellamy2013network}
\bibfield{author}{\bibinfo{person}{Marcus~A Bellamy} {and}
  \bibinfo{person}{Rahul~C Basole}.} \bibinfo{year}{2013}\natexlab{}.
\newblock \showarticletitle{Network analysis of supply chain systems: A
  systematic review and future research}.
\newblock \bibinfo{journal}{\emph{Systems Engineering}} \bibinfo{volume}{16},
  \bibinfo{number}{2} (\bibinfo{year}{2013}), \bibinfo{pages}{235--249}.
\newblock


\bibitem[\protect\citeauthoryear{Bloch, Jackson, and Tebaldi}{Bloch
  et~al\mbox{.}}{2019}]%
        {bloch2019centrality}
\bibfield{author}{\bibinfo{person}{Francis Bloch}, \bibinfo{person}{Matthew~O
  Jackson}, {and} \bibinfo{person}{Pietro Tebaldi}.}
  \bibinfo{year}{2019}\natexlab{}.
\newblock \showarticletitle{Centrality measures in networks}.
\newblock \bibinfo{journal}{\emph{Available at SSRN 2749124}}
  (\bibinfo{year}{2019}).
\newblock


\bibitem[\protect\citeauthoryear{Borrega, Taul{\'e}, and Mart{\i}}{Borrega
  et~al\mbox{.}}{2007}]%
        {borrega2007we}
\bibfield{author}{\bibinfo{person}{Oriol Borrega}, \bibinfo{person}{Mariona
  Taul{\'e}}, {and} \bibinfo{person}{M~Ant{\o}’nia Mart{\i}}.}
  \bibinfo{year}{2007}\natexlab{}.
\newblock \showarticletitle{What do we mean when we speak about Named
  Entities}. In \bibinfo{booktitle}{\emph{Proceedings of Corpus Linguistics}}.
\newblock


\bibitem[\protect\citeauthoryear{Brewitt-Taylor and Richard}{Brewitt-Taylor and
  Richard}{1931}]%
        {brewitt1931romance}
\bibfield{author}{\bibinfo{person}{Charles~Henry Brewitt-Taylor} {and}
  \bibinfo{person}{Timothy Richard}.} \bibinfo{year}{1931}\natexlab{}.
\newblock \bibinfo{booktitle}{\emph{Romance of the Three Kingdom}}.
\newblock


\bibitem[\protect\citeauthoryear{Burt}{Burt}{2009}]%
        {burt2009structural}
\bibfield{author}{\bibinfo{person}{Ronald~S Burt}.}
  \bibinfo{year}{2009}\natexlab{}.
\newblock \bibinfo{booktitle}{\emph{Structural holes: The social structure of
  competition}}.
\newblock \bibinfo{publisher}{Harvard university press}.
\newblock


\bibitem[\protect\citeauthoryear{Cardillo, G{\'o}mez-Gardenes, Zanin, Romance,
  Papo, Del~Pozo, and Boccaletti}{Cardillo et~al\mbox{.}}{2013}]%
        {cardillo2013emergence}
\bibfield{author}{\bibinfo{person}{Alessio Cardillo},
  \bibinfo{person}{Jes{\'u}s G{\'o}mez-Gardenes}, \bibinfo{person}{Massimiliano
  Zanin}, \bibinfo{person}{Miguel Romance}, \bibinfo{person}{David Papo},
  \bibinfo{person}{Francisco Del~Pozo}, {and} \bibinfo{person}{Stefano
  Boccaletti}.} \bibinfo{year}{2013}\natexlab{}.
\newblock \showarticletitle{Emergence of network features from multiplexity}.
\newblock \bibinfo{journal}{\emph{Scientific reports}} \bibinfo{volume}{3},
  \bibinfo{number}{1} (\bibinfo{year}{2013}), \bibinfo{pages}{1--6}.
\newblock


\bibitem[\protect\citeauthoryear{Chakraborti and Patriarca}{Chakraborti and
  Patriarca}{2009}]%
        {chakraborti2009variational}
\bibfield{author}{\bibinfo{person}{Anirban Chakraborti} {and}
  \bibinfo{person}{Marco Patriarca}.} \bibinfo{year}{2009}\natexlab{}.
\newblock \showarticletitle{Variational principle for the Pareto power law}.
\newblock \bibinfo{journal}{\emph{Physical review letters}}
  \bibinfo{volume}{103}, \bibinfo{number}{22} (\bibinfo{year}{2009}),
  \bibinfo{pages}{228701}.
\newblock


\bibitem[\protect\citeauthoryear{Chowdhury, Muhuri, Chakraborty, and
  Chakraborty}{Chowdhury et~al\mbox{.}}{2019}]%
        {chowdhury2019analysis}
\bibfield{author}{\bibinfo{person}{Tapan Chowdhury}, \bibinfo{person}{Samya
  Muhuri}, \bibinfo{person}{Susanta Chakraborty}, {and}
  \bibinfo{person}{Sabitri~Nanda Chakraborty}.}
  \bibinfo{year}{2019}\natexlab{}.
\newblock \showarticletitle{Analysis of adapted films and stories based on
  social network}.
\newblock \bibinfo{journal}{\emph{IEEE Transactions on Computational Social
  Systems}} \bibinfo{volume}{6}, \bibinfo{number}{5} (\bibinfo{year}{2019}),
  \bibinfo{pages}{858--869}.
\newblock


\bibitem[\protect\citeauthoryear{Devlin, Chang, Lee, and Toutanova}{Devlin
  et~al\mbox{.}}{2018}]%
        {devlin2018bert}
\bibfield{author}{\bibinfo{person}{Jacob Devlin}, \bibinfo{person}{Ming-Wei
  Chang}, \bibinfo{person}{Kenton Lee}, {and} \bibinfo{person}{Kristina
  Toutanova}.} \bibinfo{year}{2018}\natexlab{}.
\newblock \showarticletitle{Bert: Pre-training of deep bidirectional
  transformers for language understanding}.
\newblock \bibinfo{journal}{\emph{arXiv preprint arXiv:1810.04805}}
  (\bibinfo{year}{2018}).
\newblock


\bibitem[\protect\citeauthoryear{Dobrow, Chandler, Murphy, and Kram}{Dobrow
  et~al\mbox{.}}{2012}]%
        {dobrow2012review}
\bibfield{author}{\bibinfo{person}{Shoshana~R Dobrow}, \bibinfo{person}{Dawn~E
  Chandler}, \bibinfo{person}{Wendy~M Murphy}, {and} \bibinfo{person}{Kathy~E
  Kram}.} \bibinfo{year}{2012}\natexlab{}.
\newblock \showarticletitle{A review of developmental networks: Incorporating a
  mutuality perspective}.
\newblock \bibinfo{journal}{\emph{Journal of Management}} \bibinfo{volume}{38},
  \bibinfo{number}{1} (\bibinfo{year}{2012}), \bibinfo{pages}{210--242}.
\newblock


\bibitem[\protect\citeauthoryear{Elson, Dames, and McKeown}{Elson
  et~al\mbox{.}}{2010}]%
        {elson2010extracting}
\bibfield{author}{\bibinfo{person}{David Elson}, \bibinfo{person}{Nicholas
  Dames}, {and} \bibinfo{person}{Kathleen McKeown}.}
  \bibinfo{year}{2010}\natexlab{}.
\newblock \showarticletitle{Extracting social networks from literary fiction}.
  In \bibinfo{booktitle}{\emph{Proceedings of the 48th annual meeting of the
  association for computational linguistics}}. \bibinfo{pages}{138--147}.
\newblock


\bibitem[\protect\citeauthoryear{Flynn, Reagans, and Guillory}{Flynn
  et~al\mbox{.}}{2010}]%
        {flynn2010you}
\bibfield{author}{\bibinfo{person}{Francis~J Flynn}, \bibinfo{person}{Ray~E
  Reagans}, {and} \bibinfo{person}{Lucia Guillory}.}
  \bibinfo{year}{2010}\natexlab{}.
\newblock \showarticletitle{Do you two know each other? Transitivity,
  homophily, and the need for (network) closure.}
\newblock \bibinfo{journal}{\emph{Journal of personality and social
  psychology}} \bibinfo{volume}{99}, \bibinfo{number}{5}
  (\bibinfo{year}{2010}), \bibinfo{pages}{855}.
\newblock


\bibitem[\protect\citeauthoryear{Franzosi}{Franzosi}{2010}]%
        {franzosi2010quantitative}
\bibfield{author}{\bibinfo{person}{Roberto Franzosi}.}
  \bibinfo{year}{2010}\natexlab{}.
\newblock \bibinfo{booktitle}{\emph{Quantitative narrative analysis}}.
\newblock Number 162. \bibinfo{publisher}{Sage}.
\newblock


\bibitem[\protect\citeauthoryear{Kayastha, Niyato, Wang, and Hossain}{Kayastha
  et~al\mbox{.}}{2011}]%
        {kayastha2011applications}
\bibfield{author}{\bibinfo{person}{Nipendra Kayastha}, \bibinfo{person}{Dusit
  Niyato}, \bibinfo{person}{Ping Wang}, {and} \bibinfo{person}{Ekram Hossain}.}
  \bibinfo{year}{2011}\natexlab{}.
\newblock \showarticletitle{Applications, architectures, and protocol design
  issues for mobile social networks: A survey}.
\newblock \bibinfo{journal}{\emph{Proc. IEEE}} \bibinfo{volume}{99},
  \bibinfo{number}{12} (\bibinfo{year}{2011}), \bibinfo{pages}{2130--2158}.
\newblock


\bibitem[\protect\citeauthoryear{Kirschenbaum}{Kirschenbaum}{2016}]%
        {kirschenbaum2016digital}
\bibfield{author}{\bibinfo{person}{Matthew~G Kirschenbaum}.}
  \bibinfo{year}{2016}\natexlab{}.
\newblock \showarticletitle{What is digital humanities and what’s it doing in
  English departments?}
\newblock In \bibinfo{booktitle}{\emph{Defining Digital Humanities}}.
  \bibinfo{publisher}{Routledge}, \bibinfo{pages}{211--220}.
\newblock


\bibitem[\protect\citeauthoryear{M{\`a}rquez and Rodr{\'\i}guez}{M{\`a}rquez
  and Rodr{\'\i}guez}{1998}]%
        {marquez1998part}
\bibfield{author}{\bibinfo{person}{Llu{\'\i}s M{\`a}rquez} {and}
  \bibinfo{person}{Horacio Rodr{\'\i}guez}.} \bibinfo{year}{1998}\natexlab{}.
\newblock \showarticletitle{Part-of-speech tagging using decision trees}. In
  \bibinfo{booktitle}{\emph{European Conference on Machine Learning}}.
  Springer, \bibinfo{pages}{25--36}.
\newblock


\bibitem[\protect\citeauthoryear{Marrero, Urbano, S{\'a}nchez-Cuadrado, Morato,
  and G{\'o}mez-Berb{\'\i}s}{Marrero et~al\mbox{.}}{2013}]%
        {marrero2013named}
\bibfield{author}{\bibinfo{person}{M{\'o}nica Marrero},
  \bibinfo{person}{Juli{\'a}n Urbano}, \bibinfo{person}{Sonia
  S{\'a}nchez-Cuadrado}, \bibinfo{person}{Jorge Morato}, {and}
  \bibinfo{person}{Juan~Miguel G{\'o}mez-Berb{\'\i}s}.}
  \bibinfo{year}{2013}\natexlab{}.
\newblock \showarticletitle{Named entity recognition: fallacies, challenges and
  opportunities}.
\newblock \bibinfo{journal}{\emph{Computer Standards \& Interfaces}}
  \bibinfo{volume}{35}, \bibinfo{number}{5} (\bibinfo{year}{2013}),
  \bibinfo{pages}{482--489}.
\newblock


\bibitem[\protect\citeauthoryear{McPherson, Smith-Lovin, and Cook}{McPherson
  et~al\mbox{.}}{2001}]%
        {mcpherson2001birds}
\bibfield{author}{\bibinfo{person}{Miller McPherson}, \bibinfo{person}{Lynn
  Smith-Lovin}, {and} \bibinfo{person}{James~M Cook}.}
  \bibinfo{year}{2001}\natexlab{}.
\newblock \showarticletitle{Birds of a feather: Homophily in social networks}.
\newblock \bibinfo{journal}{\emph{Annual review of sociology}}
  \bibinfo{volume}{27}, \bibinfo{number}{1} (\bibinfo{year}{2001}),
  \bibinfo{pages}{415--444}.
\newblock


\bibitem[\protect\citeauthoryear{Mikolov, Chen, Corrado, and Dean}{Mikolov
  et~al\mbox{.}}{2013}]%
        {mikolov2013efficient}
\bibfield{author}{\bibinfo{person}{Tomas Mikolov}, \bibinfo{person}{Kai Chen},
  \bibinfo{person}{Greg Corrado}, {and} \bibinfo{person}{Jeffrey Dean}.}
  \bibinfo{year}{2013}\natexlab{}.
\newblock \showarticletitle{Efficient estimation of word representations in
  vector space}.
\newblock \bibinfo{journal}{\emph{arXiv preprint arXiv:1301.3781}}
  (\bibinfo{year}{2013}).
\newblock


\bibitem[\protect\citeauthoryear{Mitri}{Mitri}{2020}]%
        {mitri2020story}
\bibfield{author}{\bibinfo{person}{Michel Mitri}.}
  \bibinfo{year}{2020}\natexlab{}.
\newblock \showarticletitle{Story Analysis Using Natural Language Processing
  and Interactive Dashboards}.
\newblock \bibinfo{journal}{\emph{Journal of Computer Information Systems}}
  (\bibinfo{year}{2020}), \bibinfo{pages}{1--11}.
\newblock


\bibitem[\protect\citeauthoryear{Neal}{Neal}{2017}]%
        {neal2017small}
\bibfield{author}{\bibinfo{person}{Zachary~P Neal}.}
  \bibinfo{year}{2017}\natexlab{}.
\newblock \showarticletitle{How small is it? Comparing indices of small
  worldliness}.
\newblock \bibinfo{journal}{\emph{Network Science}} \bibinfo{volume}{5},
  \bibinfo{number}{1} (\bibinfo{year}{2017}), \bibinfo{pages}{30--44}.
\newblock


\bibitem[\protect\citeauthoryear{Otte and Rousseau}{Otte and Rousseau}{2002}]%
        {otte2002social}
\bibfield{author}{\bibinfo{person}{Evelien Otte} {and} \bibinfo{person}{Ronald
  Rousseau}.} \bibinfo{year}{2002}\natexlab{}.
\newblock \showarticletitle{Social network analysis: a powerful strategy, also
  for the information sciences}.
\newblock \bibinfo{journal}{\emph{Journal of information Science}}
  \bibinfo{volume}{28}, \bibinfo{number}{6} (\bibinfo{year}{2002}),
  \bibinfo{pages}{441--453}.
\newblock


\bibitem[\protect\citeauthoryear{Pakray, Pal, Majumder, and Gelbukh}{Pakray
  et~al\mbox{.}}{2015}]%
        {Pakray2015ResourceBA}
\bibfield{author}{\bibinfo{person}{Partha Pakray}, \bibinfo{person}{Arunagshu
  Pal}, \bibinfo{person}{Goutam Majumder}, {and} \bibinfo{person}{Alexander
  Gelbukh}.} \bibinfo{year}{2015}\natexlab{}.
\newblock \showarticletitle{Resource Building and Parts-of-Speech (POS) Tagging
  for the Mizo Language}.
\newblock \bibinfo{journal}{\emph{2015 Fourteenth Mexican International
  Conference on Artificial Intelligence (MICAI)}} (\bibinfo{year}{2015}),
  \bibinfo{pages}{3--7}.
\newblock


\bibitem[\protect\citeauthoryear{Pepperberg}{Pepperberg}{1999}]%
        {pepperberg1999rethinking}
\bibfield{author}{\bibinfo{person}{Irene~M Pepperberg}.}
  \bibinfo{year}{1999}\natexlab{}.
\newblock \showarticletitle{Rethinking syntax: A commentary on E. Kako’s
  “Elements of syntax in the systems of three language-trained animals”}.
\newblock \bibinfo{journal}{\emph{Animal Learning \& Behavior}}
  \bibinfo{volume}{27}, \bibinfo{number}{1} (\bibinfo{year}{1999}),
  \bibinfo{pages}{15--17}.
\newblock


\bibitem[\protect\citeauthoryear{Peters, Neumann, Iyyer, Gardner, Clark, Lee,
  and Zettlemoyer}{Peters et~al\mbox{.}}{2018}]%
        {peters2018deep}
\bibfield{author}{\bibinfo{person}{Matthew~E Peters}, \bibinfo{person}{Mark
  Neumann}, \bibinfo{person}{Mohit Iyyer}, \bibinfo{person}{Matt Gardner},
  \bibinfo{person}{Christopher Clark}, \bibinfo{person}{Kenton Lee}, {and}
  \bibinfo{person}{Luke Zettlemoyer}.} \bibinfo{year}{2018}\natexlab{}.
\newblock \showarticletitle{Deep contextualized word representations}.
\newblock \bibinfo{journal}{\emph{arXiv preprint arXiv:1802.05365}}
  (\bibinfo{year}{2018}).
\newblock


\bibitem[\protect\citeauthoryear{Polo, Iacono, Fiorentino, and Pierri}{Polo
  et~al\mbox{.}}{2019}]%
        {polo2019social}
\bibfield{author}{\bibinfo{person}{Maria Polo}, \bibinfo{person}{Umberto~Dello
  Iacono}, \bibinfo{person}{Giuseppe Fiorentino}, {and} \bibinfo{person}{Anna
  Pierri}.} \bibinfo{year}{2019}\natexlab{}.
\newblock \showarticletitle{A social network analysis approach to a digital
  interactive storytelling in mathematics}.
\newblock \bibinfo{journal}{\emph{Journal of e-Learning and Knowledge Society}}
  \bibinfo{volume}{15}, \bibinfo{number}{3} (\bibinfo{year}{2019}),
  \bibinfo{pages}{239--250}.
\newblock


\bibitem[\protect\citeauthoryear{Radford, Narasimhan, Salimans, and
  Sutskever}{Radford et~al\mbox{.}}{2018}]%
        {radford2018improving}
\bibfield{author}{\bibinfo{person}{Alec Radford}, \bibinfo{person}{Karthik
  Narasimhan}, \bibinfo{person}{Tim Salimans}, {and} \bibinfo{person}{Ilya
  Sutskever}.} \bibinfo{year}{2018}\natexlab{}.
\newblock \showarticletitle{Improving language understanding with unsupervised
  learning}.
\newblock \bibinfo{journal}{\emph{Technical report, OpenAI}}
  (\bibinfo{year}{2018}).
\newblock


\bibitem[\protect\citeauthoryear{Rajpurkar, Zhang, Lopyrev, and
  Liang}{Rajpurkar et~al\mbox{.}}{2016}]%
        {rajpurkar2016squad}
\bibfield{author}{\bibinfo{person}{Pranav Rajpurkar}, \bibinfo{person}{Jian
  Zhang}, \bibinfo{person}{Konstantin Lopyrev}, {and} \bibinfo{person}{Percy
  Liang}.} \bibinfo{year}{2016}\natexlab{}.
\newblock \showarticletitle{Squad: 100,000+ questions for machine comprehension
  of text}.
\newblock \bibinfo{journal}{\emph{arXiv preprint arXiv:1606.05250}}
  (\bibinfo{year}{2016}).
\newblock


\bibitem[\protect\citeauthoryear{Rosati}{Rosati}{2014}]%
        {rosati2014google}
\bibfield{author}{\bibinfo{person}{Eleonora Rosati}.}
  \bibinfo{year}{2014}\natexlab{}.
\newblock \showarticletitle{Google Books' Library Project is fair use}.
\newblock \bibinfo{journal}{\emph{Journal of Intellectual Property Law \&
  Practice}} \bibinfo{volume}{9}, \bibinfo{number}{2} (\bibinfo{year}{2014}),
  \bibinfo{pages}{104--106}.
\newblock


\bibitem[\protect\citeauthoryear{Sang and De~Meulder}{Sang and
  De~Meulder}{2003}]%
        {sang2003introduction}
\bibfield{author}{\bibinfo{person}{Erik~F Sang} {and} \bibinfo{person}{Fien
  De~Meulder}.} \bibinfo{year}{2003}\natexlab{}.
\newblock \showarticletitle{Introduction to the CoNLL-2003 shared task:
  Language-independent named entity recognition}.
\newblock \bibinfo{journal}{\emph{arXiv preprint cs/0306050}}
  (\bibinfo{year}{2003}).
\newblock


\bibitem[\protect\citeauthoryear{Strehovec}{Strehovec}{2016}]%
        {strehovec2016text}
\bibfield{author}{\bibinfo{person}{Janez Strehovec}.}
  \bibinfo{year}{2016}\natexlab{}.
\newblock \bibinfo{booktitle}{\emph{Text as Ride: Electronic Literature and New
  Media Art}}.
\newblock \bibinfo{publisher}{Center for Literary Computing}.
\newblock


\bibitem[\protect\citeauthoryear{Trieu, Tran, Ittoo, and Nguyen}{Trieu
  et~al\mbox{.}}{2019}]%
        {trieu2019leveraging}
\bibfield{author}{\bibinfo{person}{Hai-Long Trieu}, \bibinfo{person}{Duc-Vu
  Tran}, \bibinfo{person}{Ashwin Ittoo}, {and} \bibinfo{person}{Le-Minh
  Nguyen}.} \bibinfo{year}{2019}\natexlab{}.
\newblock \showarticletitle{Leveraging Additional Resources for Improving
  Statistical Machine Translation on Asian Low-Resource Languages}.
\newblock \bibinfo{journal}{\emph{ACM Transactions on Asian and Low-Resource
  Language Information Processing (TALLIP)}} \bibinfo{volume}{18},
  \bibinfo{number}{3} (\bibinfo{year}{2019}), \bibinfo{pages}{1--22}.
\newblock


\bibitem[\protect\citeauthoryear{Vaswani, Shazeer, Parmar, Uszkoreit, Jones,
  Gomez, Kaiser, and Polosukhin}{Vaswani et~al\mbox{.}}{2017}]%
        {vaswani2017attention}
\bibfield{author}{\bibinfo{person}{Ashish Vaswani}, \bibinfo{person}{Noam
  Shazeer}, \bibinfo{person}{Niki Parmar}, \bibinfo{person}{Jakob Uszkoreit},
  \bibinfo{person}{Llion Jones}, \bibinfo{person}{Aidan~N Gomez},
  \bibinfo{person}{{\L}ukasz Kaiser}, {and} \bibinfo{person}{Illia
  Polosukhin}.} \bibinfo{year}{2017}\natexlab{}.
\newblock \showarticletitle{Attention is all you need}. In
  \bibinfo{booktitle}{\emph{Advances in neural information processing
  systems}}. \bibinfo{pages}{5998--6008}.
\newblock


\bibitem[\protect\citeauthoryear{Waumans, Nicod{\`e}me, and Bersini}{Waumans
  et~al\mbox{.}}{2015}]%
        {waumans2015topology}
\bibfield{author}{\bibinfo{person}{Micha{\"e}l~C Waumans},
  \bibinfo{person}{Thibaut Nicod{\`e}me}, {and} \bibinfo{person}{Hugues
  Bersini}.} \bibinfo{year}{2015}\natexlab{}.
\newblock \showarticletitle{Topology analysis of social networks extracted from
  literature}.
\newblock \bibinfo{journal}{\emph{PloS one}} \bibinfo{volume}{10},
  \bibinfo{number}{6} (\bibinfo{year}{2015}), \bibinfo{pages}{e0126470}.
\newblock


\bibitem[\protect\citeauthoryear{Worth}{Worth}{2008}]%
        {worth2008storytelling}
\bibfield{author}{\bibinfo{person}{Sarah~E Worth}.}
  \bibinfo{year}{2008}\natexlab{}.
\newblock \showarticletitle{Storytelling and narrative knowing: An examination
  of the epistemic benefits of well-told stories}.
\newblock \bibinfo{journal}{\emph{Journal of Aesthetic Education}}
  \bibinfo{volume}{42}, \bibinfo{number}{3} (\bibinfo{year}{2008}),
  \bibinfo{pages}{42--56}.
\newblock


\bibitem[\protect\citeauthoryear{Zen and Sak}{Zen and Sak}{2015}]%
        {zen2015unidirectional}
\bibfield{author}{\bibinfo{person}{Heiga Zen} {and}
  \bibinfo{person}{Ha{\c{s}}im Sak}.} \bibinfo{year}{2015}\natexlab{}.
\newblock \showarticletitle{Unidirectional long short-term memory recurrent
  neural network with recurrent output layer for low-latency speech synthesis}.
  In \bibinfo{booktitle}{\emph{2015 IEEE International Conference on Acoustics,
  Speech and Signal Processing (ICASSP)}}. IEEE, \bibinfo{pages}{4470--4474}.
\newblock


\bibitem[\protect\citeauthoryear{Zhang, Zhao, Xu, and Wang}{Zhang
  et~al\mbox{.}}{2014}]%
        {zhang2014small}
\bibfield{author}{\bibinfo{person}{Jun Zhang}, \bibinfo{person}{Hai Zhao},
  \bibinfo{person}{Jiu-qiang Xu}, {and} \bibinfo{person}{Jin-fa Wang}.}
  \bibinfo{year}{2014}\natexlab{}.
\newblock \showarticletitle{Small-world and Scale-free Features in Harry
  Potter}.
\newblock  (\bibinfo{year}{2014}).
\newblock


\bibitem[\protect\citeauthoryear{Zhou and Mondrag{\'o}n}{Zhou and
  Mondrag{\'o}n}{2004}]%
        {zhou2004rich}
\bibfield{author}{\bibinfo{person}{Shi Zhou} {and} \bibinfo{person}{Ra{\'u}l~J
  Mondrag{\'o}n}.} \bibinfo{year}{2004}\natexlab{}.
\newblock \showarticletitle{The rich-club phenomenon in the Internet topology}.
\newblock \bibinfo{journal}{\emph{IEEE Communications Letters}}
  \bibinfo{volume}{8}, \bibinfo{number}{3} (\bibinfo{year}{2004}),
  \bibinfo{pages}{180--182}.
\newblock


\end{thebibliography}

\appendix









\end{document}